\documentclass[useAMS,usenatbib]{mn2e}

\newcommand{\ie}{i.e.~}
\newcommand{\eg}{e.g.~}
\newcommand{\pms}{P_{\mathrm{MS}}}     	  		   
\newcommand{\pwd}{P_{\mathrm{WD}}}	  		   
\newcommand{\pwdh}{P_{\mathrm{WD}}^{\mathrm{H}}}  	   
\newcommand{\pwdhe}{P_{\mathrm{WD}}^{\mathrm{He}}} 	   
\newcommand{\mwd}{M_{\mathrm{WD}}}	  		   
\newcommand{\mms}{M_{\mathrm{MS}}}	  		   
\newcommand{\msolar}{M_{\odot}}		  		   
\newcommand{\mbol}{M_{\mathrm{bol}}}	  		   
\newcommand{\tform}{t}		  	  		   
\newcommand{\tms}{t_{\mathrm{MS}}}	  		   
\newcommand{\tcool}{t_{\mathrm{cool}}}    		   
\newcommand{\tmax}{T_{\mathrm{0}}}	  		   
\newcommand{\ifmr}{\xi}			  		   
\newcommand{\imf}{\phi}		  	  		   
\newcommand{\sfr}{\psi}		  	  		   
\newcommand{\wdlf}{\Phi}		  		   
\newcommand{\msmassfnt}{\mms^{\mathrm{lifetime}}(\tform)}  

\usepackage[cmex10]{amsmath}
\usepackage{natbib}
\usepackage{graphicx}
\usepackage{subfigure}
\usepackage{amssymb}

\title[Solar Neighbourhood Star Formation History from the WDLF]{The Star Formation History of the Solar Neighbourhood from the White Dwarf Luminosity Function.}
\author[N. Rowell]{N. Rowell\thanks{E-mail: nickrowell@computing.dundee.ac.uk}\\
Space Technology Centre, School of Computing, University of Dundee}
\begin{document}

\date{Accepted xx.xx.xxxx. Received xx.xx.xxxx; in original form xx.xx.xxxx}

\pagerange{\pageref{firstpage}--\pageref{lastpage}} \pubyear{2002}

\maketitle

\begin{abstract}
The termination in the white dwarf luminosity function is a standard diagnostic tool for measuring the total age of nearby stellar populations.
In this paper, an algorithm is presented for inverting the full white dwarf luminosity function to obtain a maximum likelihood estimate
of the time varying star formation rate of the host stellar population.
%
Tests with synthetic data demonstrate that the algorithm converges over a wide class of underlying star formation rate forms.
The algorithm successfully estimates the moving average star formation rate as a function of lookback time in the presence of
realistic measurement noise, though suffers from degeneracies around discontinuities in the underlying star formation rate.
The inversion results are most sensitive to the choice of white dwarf cooling models, with the models
produced by different groups giving quite different results. The results are relatively insensitive to the progenitor metallicity, 
initial mass function, initial-final mass relation and ratio of H/He atmosphere white dwarfs.
Application to two independent determinations of the Solar neighbourhood white dwarf luminosity function
gives similar results. The star formation rate has a bimodal form, with broad peaks at 2--3 Gyr and 7--9 Gyr in the past,
separated by a significant lull of magnitude 30--90\% depending on choice of cooling models. The onset of star formation occurs around 8--10 Gyr ago.
The total integrated star formation rate is~$\sim0.014$~stars/pc$^{3}$ in the Solar neighbourhood, for stars more massive than $0.6\msolar$.
\end{abstract}
\begin{keywords}
Solar neighbourhood; white dwarfs; Galaxy: formation
\end{keywords}

\section{Introduction}
The white dwarf luminosity function (WDLF) is a useful tool for determining the age of a population of stars.
The magnitude at which the function terminates is time dependent, and by fitting the faint end with theoretical WDLF models
of different ages one can obtain a statistical estimate of the age of the population without having to
determine the total age of any individual white dwarf, which is considerably more difficult.
%
%
This technique has been applied successfully to single burst populations such as open and globular clusters 
(\eg NGC 2158 \citep{bedin2010}; M4 \citep{bedin2009}), where the comparison of WD and MS ages has proved to be very fruitful.
NGC 6791 in particular has provided an important benchmark for understanding WD cooling processes at faint magnitudes
\citep{Garcia-Berro2010} and the binary fractions of old, metal rich clusters \citep{bedin2008}.

The Galactic disk WDLF has been examined many times over the years \citep{winget1987,oswalt1996,leggett1998}, 
with studies finding ages in the range 8-10 Gyr depending on the WD evolutionary models adopted.
For continuous populations such as the disk, the faintest WDs
are the descendents of high mass MS stars that formed at very early times, and their lifetimes are completely dominated
by the WD cooling phase, leading to age estimates largely independent of uncertainties associated with MS lifetimes.
The picture is considerably more complicated at brighter magnitudes, because the WDs are a mixture of ages: both young, high mass WDs that are
produced by recently-formed MS progenitors, and old, low mass WDs produced by low mass MS stars that formed at early times.
It is for this reason that nearly all studies have focussed exclusively on the faint turnover of the disk WDLF in an attempt to constrain
the total disk age.
%
%
%

The groundbreaking study of \citet{noh1990} revealed that the detailed shape of the WDLF at magnitudes brighter than the
peak is far more sensitive to the time varying star formation rate (SFR) than to variations in the initial mass function (IMF).
By forward modelling methods, they demonstrated that a marginal feature in the WDLF at $\mbol \approx 10$
could be interpreted as evidence for a burst of star formation $0.3$ Gyr ago.
According to both \citet{iben1989} and \citet{noh1990}, the shape of the WDLF at intermediate magnitudes is also affected by the
cooling rates of WDs: magnitudes at which the cooling is faster tend to have lower WD numbers, due to WDs transiting quickly to fainter
magnitudes.
Other authors have used this fact to examine the WDLF for evidence of additional WD cooling mechanisms, beyond those currently
included in WD cooling theory. One possibility is cooling by emission of `axions' \citep[\eg][]{isern2008, melendez2012}, a 
light pseudoscalar particle postulated by the Peccei-Quinn theory.
\citet{isern2008} argue that the rising slope of the WDLF at $\mbol<13$ is independent of age, with the shape determined solely by
WD cooling rates. By forward modelling the WDLF assuming a constant SFR, and with WD cooling models that include axion emission,
they find a best fit axion mass of 5meV.
\subsection{Forward modelling the WDLF}
The standard equation for modelling the WDLF for a given star formation history is \citep[\eg][]{iben1989,fontaine2001}
\begin{equation}
\wdlf(\mbol) = \int\limits^{M_u}\limits_{M_l} \frac{d\tcool}{d\mbol}
\, \sfr(\tmax - \tcool - \tms) \, \imf(M) \, dM      
\label{eq:theoryLF}                             
\end{equation}
where $\wdlf(\mbol)$ is the number density of WDs at magnitude $\mbol$.
The derivative inside the integral is the characteristic cooling time for WDs, $\sfr(t)$ is the star formation rate at time $t$
and $\imf$ is the initial mass function. The integral also depends on the lifetimes of main sequence progenitors as a function
of mass and metallicity $\tms$, the WD cooling times as a function of mass and luminosity $\tcool$, and the total
time since the onset of star formation $\tmax$. The integral is over all main sequence masses that have had time to produce WDs at the present
day, with the magnitude-dependent lower limit corresponding to the solution of
\begin{equation}
\tmax = \tcool(\mbol,\ifmr(M_l)) + \tms(M_l,Z)
\label{eq:lowerLimit}
\end{equation}
and the upper limit for WD production $M_u \approx 7$ M$_{\odot}$. $\ifmr$ is the initial-final mass relation (IFMR) that relates the
mass of a MS star to the mass of the WD that it forms.
We note in passing that for modelling single burst populations, inserting a delta function for the SFR simplifies equation \ref{eq:theoryLF}
to
\begin{equation}
\wdlf(\mbol) =  \imf(M) \frac{dM}{d\mbol}.
\label{eq:singleBurst}
\end{equation}
This can also be derived by considering the conservation of stars between corresponding progenitor mass and WD bolometric magnitude
intervals, $\wdlf(\mbol) d\mbol = \imf(M) dM$. For such populations, there is a one-to-one correspondence between WD bolometric
magnitude and MS progenitor mass, given by the solution (if any) to equation \ref{eq:lowerLimit}.
\subsection{On the invertibility of the WDLF}
Although there have been several major studies to develop forward modelling approaches to estimating the Galactic age from 
the WDLF \citep[\eg][]{winget1987,oswalt1996,leggett1998},
it appears that not much work has been done on the possibility of \textit{inverting} the WDLF to obtain a direct estimate of the SFR.
This is in stark contrast to comparable studies using MS stars, for which very mature Bayesian methods have been developed for inverting
colour-magnitude diagrams (CMDs). \citet{hernandez2000} and \citet{vergely2002} develop largely independent Bayesian techniques for 
inverting the Hipparcos CMD to measure the Solar neighbourhood SFR. The non-parametric nature of these approaches is a great benefit of the
inversion method: forward modelling techniques normally involve selecting a parameterisation for the SFR then optimizing the parameters
using some choice of cost function to compare models to the data (see, for example, \citet{bertelli2001}). The solution is therefore 
only optimal in the context of the
adopted model, which might not correspond to reality. Non-parametric methods allow the full form of the SFR solution to be determined from the
data alone, rather than forcing it to conform to some imposed function.

It has generally been thought that the shape of the WDLF is almost independent of the SFR at all but the faintest magnitudes,
where it is governed mainly by the total population age. \citet{isern2008} 
present the following argument for the \emph{bright} portion of the WDLF being independent of the SFR:
first, the characteristic cooling time of a WD is not very sensitive to mass so the derivative in equation  \ref{eq:theoryLF}
can be taken out of the integral and replaced with an average over all WD masses.
Because the cooling rates of WDs are highly non-linear,
at bright magnitudes the cooling time is relatively small (on the order of 200 Myr at $\mbol=10$) and the lower limit of the integral 
(determined from equation \ref{eq:lowerLimit}) is satisfied by low mass stars and is almost constant. Therefore, as long as the SFR is a well-behaved
function (no large bursts or lulls within the last $\sim200$ Myr) and $\tmax$ is large enough, the integral is not sensitive to the
WD luminosity and the WDLF is determined almost entirely by the average cooling rates of WDs.

\citet{isern2012} take this further and assert that the WDLF is in fact non-invertible, due to equation \ref{eq:theoryLF} 
failing to satisfy the Picard-Lindel\"{o}f theorem for the inversion of integral equations,
meaning that the solution is sensitive to the trial function used and its uniqueness therefore not guaranteed, ultimately due 
to the smoothing effect of the integral washing out sensitivity to high frequency components of the SFR.
Some justification for this is provided by Fig. \ref{fig:wdlfs_diff_sfr} 
which shows the similarity between WDLFs computed using quite different assumptions for the star formation rate (reproduction of 
\citet{isern2012} Fig. 5). Brighter than $\mbol \sim 13$, these are almost totally degenerate. In the lower panel, the mean
age of WDs is roughly constant at these magnitudes, but fainter than $\mbol \sim 12-13$ there is a correlation between
magnitude and age. It is at these magnitudes that we expect the shape of the WDLF to contain information on time variations
in the SFR.
\begin{figure}
\centering
\includegraphics[trim=0cm 0cm 0cm 0cm, clip=true, width=8cm]{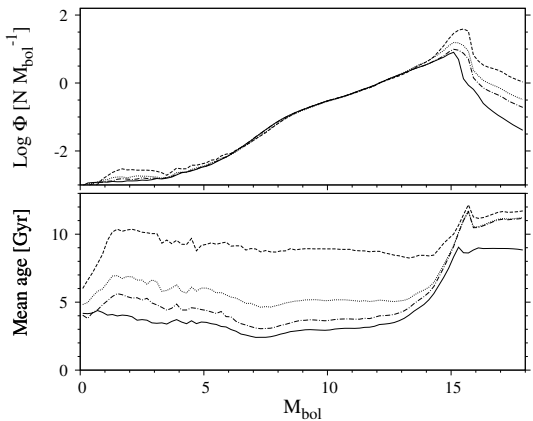}
\caption[]{
Upper panel: simulated WDLFs for a variety of different star formation histories and ages. Solid and dot-dash lines are for a constant
SFR with $\tmax=10,13$ Gyr respectively; dashed line is for $\sfr \propto \exp(-t/\tau)$ with $\tau=3$ Gyr and $\tmax=13$ Gyr; dotted line is
for $\sfr \propto (1+\exp((t-t^{\prime})/\tau)^{-1}$ with $t^{\prime}=10$ Gyr, $\tau=3$ Gyr and $\tmax=13$ Gyr.
Lower panel: mean total stellar age as a function of WD bolometric magnitude, for the same star formation histories as the upper panel.
}
\label{fig:wdlfs_diff_sfr}
\end{figure}
It's important to point out that although the star formation histories included in this figure cover a broad range of scenarios, they are all
very smooth functions with few degrees of freedom and little information content. We will demonstrate in this paper
that for such smooth star formation histories, the few WDLF points around the peak are indeed sufficient to recover the overall
form of the time varying SFR by inversion methods, and that the non-uniqueness of the solution is limited to a loss of resolution around
discontinuities.
In addition, a simple trial function (in this case a constant SFR) can recover a wide range of different 
underlying SFR forms with a good degree of confidence.
\subsection{This paper}
This paper presents the results of work to develop an inversion algorithm suitable for application to the WDLF, in order to measure the
time varying SFR.
To a first approximation, the two parameters that determine the total age of a WD are the present day bolometric magnitude, and the mass.
These can be used to determine both the total WD cooling time and the time spent on the main sequence.
The approach to inverting the WDLF developed in this paper is based on the observation that if the distribution of WD mass was known at all magnitudes,
then the WDLF could be directly transformed to the SFR, due to the correspondence between points in the progenitor mass/formation time
plane and the white dwarf mass/luminosity plane.
As this quantity is generally not known observationally, this direct approach cannot be used. 
Instead, we use an inversion technique based on the Expectation Maximization method \citep{Dempster1977,Do2008},
which is used to obtain maximum likelihood estimates of the solution to inverse problems in the presence of missing data.
This technique is used widely in image restoration, where it is called Richardson-Lucy deconvolution \citep[see also][Appendix C]{binney1998}.
Applied to the WDLF, this iterative technique involves using an initial guess of the star formation rate to derive the present day 
white dwarf mass/luminosity distribution, which is then normalised to the observed luminosity function, before transforming back 
to obtain an improved estimate of the star formation rate.

Note that although all the information pertaining to the time varying SFR lies in the region fainter than $\mbol \sim 12-13$,
in the inversions presented in this paper we will include the hot branch of the WDLF: while this doesn't help to
constrain the time varying SFR, it improves overall constraint and reduces global errors on the SFR.
Also, valid solutions should at least be consistent with this region of the WDLF so it provides an additional sanity check
on the results.

This work is motivated by two related questions: given current WD cooling models, what constraint 
can features in the WDLF at all magnitudes
place on the time varying SFR? And as a corrolary to this: can features in the WDLF be
explained exclusively by time variations in the SFR, or are additional cooling mechanisms required?
\section{Statistical Framework}
%
%
The WDLF inversion algorithm involves iteratively refining an initial guess of the SFR. 
The general procedure for each iteration is as follows. The starting point is an initial 
guess of the star formation rate $\sfr_0$,
\begin{equation}
\sfr_0  \equiv \sfr_0(\tform)
\end{equation}
with $\tform$ the lookback time, and $\sfr$ measured in units of stars per year.
In the present work $\sfr_0$ is flat, i.e. a constant star formation rate. It will be demonstrated
later that this is sufficient to recover a wide range of different underlying star formation rate forms.
This is combined with the initial mass function $\imf$ to get the joint mass and formation time distribution of main sequence progenitors
$\pms$, where
%
\begin{equation}
\pms(\mms, \tform) = \imf(\mms) \sfr_0(\tform).
\end{equation}
$\pms$ is thus separable at this stage, assuming that the initial mass function $\imf$ is independent of time.
$\pms$ is the distribution of WD progenitors, \ie the subset of main sequence stars that
form WDs at the present day. The region of the $[\mms, \tform]$ plane that these stars inhabit
is bounded by the functions
\begin{equation}
\begin{split}
\mms^{\mathrm{upper}} & = \mms^{\mathrm{max}}\\
\mms^{\mathrm{lower}} & = \msmassfnt\\
\tform_{\mathrm{upper}} & = \tmax
\end{split}
\end{equation}
where $\mms^{\mathrm{max}} = 7.0 \msolar$ is the maximum progenitor mass for WD formation.
The function $\msmassfnt$ is the mass of the main sequence star
with lifetime $\tform$; main sequence stars with lifetimes longer than the lookback time do not have
time to form WDs at the present day. $\tmax$ is the maximum lookback time, and is
a parameter of the algorithm. Note that $\tmax$ does not enforce a fixed total age
on the stellar population, because for populations younger than $\tmax$ the star formation rate
will be driven towards zero at lookback times less than $\tmax$.
Its main purpose is to exclude unphysical solutions. No \textit{a priori} knowledge of the 
total time since the onset of star formation is required. Figure \ref{fig:pms} shows an example
$\pms$ distribution generated during testing.
\begin{figure}
\centering
\includegraphics[trim=0.7cm 0cm 0.7cm 0, clip=true, width=8.5cm]{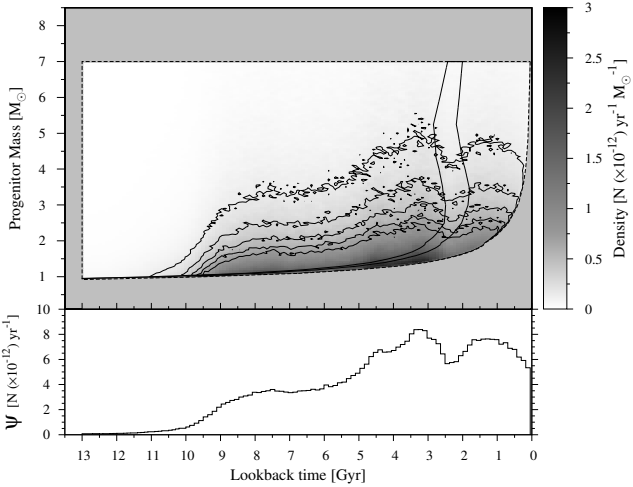}
\caption[]{Example plot of $\pms$. In the top panel, the upper grey region of the plane corresponds
to MS stars that do not produce WDs; the lower grey region corresponds to MS stars
that have not had time to produce WDs at the present day. The plane is bounded on the left
by the maximum lookback time, set to 13 Gyr here. The contours lie at intervals of $0.15\times10^{-12}$yr$^{-1}\msolar^{-1}$;
for clarity only the first six are plotted. The solid black lines mark the region inhabited by progenitors that form
$13 < \mbol < 13.5$ WDs (H atmosphere) at the present day.
The lower panel shows the star formation rate after integrating over the progenitor mass.}
\label{fig:pms}
\end{figure}

Using standard rules of probability density functions, $\pms$ is transformed to the joint mass and bolometric magnitude 
distribution of white dwarfs $\pwd(\mwd, \mbol)$ at the present day:
\begin{equation}
\pwd(\mwd, \mbol) = \pms(\mms, \tform). \left| \frac{\partial(\mms,\tform)}{\partial(\mwd,\mbol)} \right|
\end{equation}
where the Jacobian expresses the transformation of the $\mathrm{d}\mms\mathrm{d}\tform$ area element in order to conserve
the volumetric probability between corresponding intervals in $\pms$ and $\pwd$.
The $(\mwd, \mbol)$ plane is not fully populated; it is bounded at high mass by the line
\begin{equation}
\mwd^{\mathrm{upper}} = \ifmr(\mms^{\mathrm{max}})
\end{equation}
and at bright and faint magnitudes by the functions
%
%
\begin{equation}
\begin{split}
\mbol^{\mathrm{upper}} = & \mbol(\mwd, \tcool = \tmax-\tms(\ifmr^{-1}(\mwd),Z))\\
\mbol^{\mathrm{lower}} = & \mbol(\mwd, \tcool = 0)
\end{split}
\end{equation}
The WD bolometric magnitude at given mass and cooling time $\mbol(\mwd,\tcool)$ is obtained from theory. Figure \ref{fig:pwd} shows an
example $\pwd$ distribution generated during testing.
\begin{figure}
\centering
\includegraphics[trim=0.7cm 0cm 0.7cm 0, clip=true, width=8.5cm]{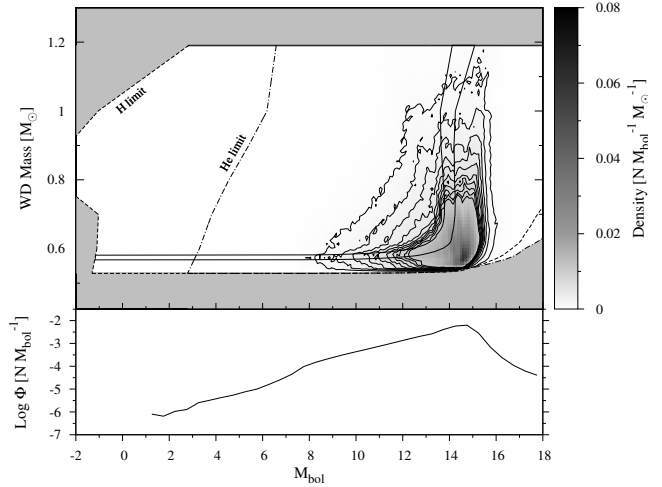}
\caption[]{Example plot of $\pwd$, transformed from the $\pms$ distribution in Fig. \ref{fig:pms}. 
Both H and He atmosphere WDs are included in this plot ($\alpha=0.5$; see section 
\ref{atmospheres} below). In the top panel, the dashed and dot-dashed lines mark the boundaries of the regions inhabited
by these objects. The upper grey region lies above the upper mass limit for WD formation and
is uninhabited. The lower grey region corresponds to WD masses that have not had time to form.
The contours lie at intervals of $0.01\mbol^{-1}\msolar^{-1}$; for clarity only the first ten are plotted.
The solid black lines mark the region inhabited by WDs (H atmosphere) of total age $3 < \tform < 4$ Gyr.
The lower panel shows the WDLF obtained by integrating over the WD mass.}
\label{fig:pwd}
\end{figure}

Because both main sequence lifetimes and white dwarf cooling rates are mass-dependent,
$\pwd$ is not separable; the variables $(\mwd, \mbol)$ are reasonably correlated, with WDs of all 
masses existing at bright magnitudes, and faint magnitudes inhabited exclusively by high mass WDs.
$\pwd$ can, however, be separated into a product of the marginal luminosity distribution $\wdlf_{\mathrm{sim}}(\mbol)$ and the
conditional probability of $\mwd$ given $\mbol$
\begin{equation}
\pwd(\mwd, \mbol) = \wdlf_{\mathrm{sim}}(\mbol) \pwd(\mwd | \mbol)
\label{eqn:rl}
\end{equation}
The quantity $\wdlf_{\mathrm{sim}}(\mbol)$ is just the WDLF for the initial guess SFR model, up to a normalisation factor.
The next crucial step is to replace the simulated WDLF in equation \ref{eqn:rl} with the observed WDLF
$\wdlf_{\mathrm{obs}}(\mbol)$, in order to obtain the updated joint distribution $\pwd^{\prime}$, where
\begin{equation}
\pwd^{\prime}(\mwd, \mbol) = \wdlf_{\mathrm{obs}}(\mbol) \pwd(\mwd | \mbol)
\label{eqn:update}
\end{equation}
This updated WD distribution has the same marginal luminosity distribution as the observed WDLF, and the magnitude-dependent mass 
distribution derived from the initial guess star formation rate model.
We can now transform this distribution to obtain the updated distribution for main sequence stars $\pms^{\prime}$
again using standard transformation rules:
\begin{equation}
\pms^{\prime}(\mms, \tform) = \pwd^{\prime}(\mwd, \mbol) \left| \frac{\partial(\mwd,\mbol)}{\partial(\mms,\tform)} \right|.
\label{eqn:trans_pms}
\end{equation}
In general $\pms^{\prime}(\mms, \tform)$ is not separable; the correction to $\pms$ produces a distribution $\pms^{\prime}$ for which 
the marginal main sequence mass distribution varies over time. It is implicit in the algorithm that the IMF is
independent of time, and that valid solutions should have this property.
It will be demonstrated emprically that as the algorithm proceeds, it converges towards a star formation rate that produces
a present day WDLF that is an increasingly better match to the observed WDLF.
Close to convergence, the correction to $\pms$ is very small, and $\pms^{\prime}$ becomes separable as required.
%

The final step is to marginalise $\pms^{\prime}$ over the main sequence mass, to obtain the updated 
star formation rate model $\sfr_1$:
\begin{equation}
\sfr_1(\tform)  = \frac{1}{1-A(\tform)} \int_{\mms^{\mathrm{lifetime}}(\tform)}^{\mms^{\mathrm{\mathrm{max}}}} \! \pms^{\prime}(\mms, \tform) \, \mathrm{d} \mms.
\label{eqn:updated_sfr}
\end{equation}
The integral is over all main sequence stars that produce WDs at the present day.
The factor $\frac{1}{1-A(\tform)}$ corrects for low mass main sequence stars that have not had time to form WDs at the present day, and
is calculated by
\begin{equation}
\label{eqn:low_mass_factor}
A(\tform) = \int_{0.6\msolar}^{\msmassfnt} \! \imf(\mms) \, \mathrm{d} \mms
\end{equation}
The lower mass limit here is chosen as $0.6 \msolar$; the IMF is still well constrained at this mass, and no star 
less massive than this forms a WD in any realistic lookback time, so using this as the normalisation point avoids
overcorrection. The star formation rate that we measure therefore only accounts for stars more massive than $0.6 \msolar$.
\subsection{Additional considerations}
%
%
\subsubsection{White dwarf atmosphere types}
\label{atmospheres}
Along with the mass and present luminosity, the H/He atmosphere type is a third parameter affecting the cooling time of a WD.
This has a significant effect at larger cooling times ($\gtrsim6$ Gyr depending on the choice of models), 
with hydrogen atmosphere WDs being brighter at a given age.
In this paper we use the cooling sequences of \citet{fontaine2001} and \citet{Salaris2010} (see Section \ref{wd_models}),
each of which provide cooling times for both H and He atmosphere WDs.

The two atmosphere types can be included in the algorithm in a relatively straightforward manner.
We compute the joint mass and bolometric magnitude distribution separately for each type to
obtain $\pwdh$ and $\pwdhe$.
Under the assumption that the atmosphere 
type is fixed at birth for WDs, these functions can be superposed to obtain the full distribution
\begin{equation}
\pwd = \alpha \pwdh + (1-\alpha) \pwdhe
\end{equation}
where $\alpha = \frac{N_{\mathrm{H}}}{N_{\mathrm{H}}+N_{\mathrm{He}}}$ lies in the range 0:1. 
Observational determinations of $\alpha$ are complicated by the fact that
the spectral energy distribution of cooling WDs appears to evolve over time, presumably due to convective
mixing of stratified layers. \citet{tremblay2008} find a value of $\frac{N_{\mathrm{He}}}{N_{\mathrm{H}}}=0.25$ to
be appropriate at high temperatures 
($T_{\mathrm{eff}}\gtrsim12,000K$), before the effect of convective mixing has set in; this
corresponds to $\alpha=0.8$, which is the value we adopt in this work.
\subsubsection{Undetected WDs and algorithm convergence}
%
Observational determinations of the WDLF generally don't cover the full range of bolometric magnitude.
%
In general due to the intrinsic rarity of very bright WDs, and the difficulty in detecting very faint WDs, there are
ranges over which the number density of WDs is unconstrained observationally. However, it is still possible to constrain
the star formation rate over the full history, because at any given formation time stars produce WDs over a relatively wide
range of $\mbol$, and as long as some of these are observed, then the star formation rate can in principle be constrained.


The implication of this with respect to the inversion algorithm is that at any given formation time, some fraction
of the WD progenitors may produce WDs that are undetected at the present day, \ie lie outside the range of the observed WDLF used to
constrain the algorithm. These `unobserved' stars will lie in one or more non-contiguous ranges of progenitor mass, corresponding
to intervals in bolometric magnitude where the WD density is unconstrained observationally.
The probability density of these objects remains at the initial value on applying
equation \ref{eqn:update}, and they provide no constraint on the star formation rate.
This can cause the algorithm to take a long time to converge, especially at times where
the true star formation rate is very low relative to the initial guess. The lower constraint
also leads to a greater risk of systematic error on the recovered star formation rate.

A better way to handle unobserved stars is to exclude them from the integral when calculating the updated star formation
rate (equation \ref{eqn:updated_sfr}), then to multiply $\sfr_1$ by a suitable factor to correct for the fraction of missing
stars. The correction factor is determined from the IMF, in the same way as for low mass stars.
This allows the algorithm to converge much faster, although the risk of systematic error remains.
For this reason, we record the fraction of unobserved stars as a function of formation time as a diagnostic for the 
systematic error on the recovered star formation rate.
In practise however, the observed WDLF--at least in the case of the Galactic disk--covers almost the entire bolometric 
magnitude range, and only a few percent of WDs at the faintest magnitudes are missing.

%
%
%
%
\subsubsection{Lookback time resolution}
The lower limit on the lookback time is set by the lifetime of the most massive WD progenitor, which in practise is 
$\sim$10$^7$--10$^8$ years depending on metallicity
\citep{Bertelli2008,Bertelli2009}.
%
%
%
The algorithm is completely insensitive
to variations in the star formation rate that occured more recently than this.
There is no theoretical upper limit on the lookback time, although most IFMRs break down at low mass 
(\eg $0.48\msolar$ for \citet{kalirai2008}, $0.47\msolar$ for \citet{catalan2008}),
producing WDs more massive than their MS progenitors. As these live for much longer than a Hubble time, this imposes no
constraint on realistic models.
%
%
%
%
%
%
Between these extremes, the finite magnitude resolution of the observational WDLF leads to a lower limit on the star formation rate time resolution.
At any age, there is a frequency above which variations in the star formation rate produce no discernible change in the observed WDLF.
Generally the resolution is poorer at older times, due to the cooling rates of WDs slowing with age. 

Considering this, the set of lookback time bins used to represent the initial guess SFR need to be selected with some care.
Attempting to match bins sizes to cooling times
of WDs over constant intervals in magnitude (so that, for example, more recent bins are narrower where cooling is faster) is attractive from
an observational point of view, but suffers from highly underpopulated bins at recent times where cooling is very rapid and
only high mass stars contribute.

A scheme using lookback time intervals corresponding to a constant number of WDs at the present day
(so that, for example, older time bins are narrower where the density of WD progenitors is greater) is attractive due to the statistical noise being
roughly uniform in each time bin, but requires very narrow magnitude bins around the peak of the WDLF which is not justified observationally  
because magnitude errors are too large.

A simple compromise between these two is to use lookback time bins of a constant width. This is the approach taken in this work.
However, it should be remembered that high frequency components of the underlying SFR are likely to be lost at older times.
Experiments with synthetic data presented in the following sections will attempt to quantify this.
\subsubsection{Convergence criteria}
The convergence of the inversion algorithm is assessed by checking the goodness of fit between the simulated and observed
WDLFs. Once a star formation rate has been arrived at that results in a WDLF that closely matches the observed
present day WDLF, no further improvement can be made and the algorithm must be halted to prevent further iterations
from over-fitting the noise in the data, causing unrealistic spikes to develop in the star formation rate.

The goodness of fit is measured using the $\chi^2$ statistic between the simulated and observed WDLFs 
\begin{equation}
\chi^2 = \sum\limits_{k} \frac{(\wdlf^k_{\mathrm{sim}} - \wdlf^k_{\mathrm{obs}})^2}{\sigma^2_{\wdlf^{k}_{\mathrm{obs}}}},
\label{eqn:chi2}
\end{equation}
where the sum is over all bolometric magnitude bins in the observed WDLF.

Tests with synthetic data suggest that convergence is reached when the relative change in $\chi^2$ from one 
iteration to the next falls below a threshold of approximately one percent. This level allows a good fit to be reached, while 
avoiding over-fitting of the data. In order to prevent statistical noise in $\chi^2$ from affecting the
convergence, we use a sliding linear fit to the most recent five $\chi^2$ values and calculate the relative
change at the latest iteration from this. 
%
%
The first iteration is always omitted from the fit, because
if the initial guess is significantly far from the truth then the first $\chi^2$ is an outlier.
%
%
%
%
%
%
\section{Monte Carlo modelling procedure}
In the present work, the inversion algorithm is implemented using a Monte Carlo method, which is described now.
The initial guess star formation rate  (in units of stars-per-year) is first broken into a fixed number of discrete 
lookback time bins of width $\delta \tform$
\[
\sfr_0(\tform)  \rightarrow \sfr_0^{\tform}
\]
over which the SFR is assumed to be constant.
A finite number $N$ of simulated stars are generated in each bin with formation times drawn uniformly within 
the range of the bin, and masses drawn from the IMF. We also randomly assign a fixed H/He WD atmosphere type to each 
star at this point: stars are assigned an H atmosphere with probability $\alpha$.
Initially, each simulated star represents $n = 1 \pm 1$ real stars, assuming Poisson statistics.
Within each bin, the number of real stars that each simulation star represents is then scaled to
\[
n = \frac{\sfr_0^{\tform} \delta\tform}{N}
\]
and the variance propagated to obtain
\[
\sigma^2_n = \left( \frac{\sfr_0^{\tform} \delta\tform}{N} \right)^2.
\]

The simulated stars are then evolved to the present day, and binned according to their WD bolometric magnitude
at a resolution that matches the observed WDLF used as input. Unobserved stars that fall outside the range of the
WDLF are identified at this point, but are not purged from the simulated population.
Prior to binning, we add a bolometric magnitude error to each star
drawn from a Gaussian distribution with width $\sigma_M$, which is designed to simulate photometric parallax errors.
The value of $\sigma_M$ should be close to the approximate size of bolometric magnitude errors on the
observed WDLF.
The luminosity function (in units of stars-per-magnitude) in a given bolometric magnitude bin $k$ is
obtained for the simulated population by
\begin{equation}
\label{eqn:sim_wdlf}
\wdlf^k_{\mathrm{sim}} = \frac{\displaystyle\sum\limits_{i=1}^{N_k} n_i^k }{\delta\mbol^k}
\end{equation}
with associated statistical uncertainty
\begin{equation}
\label{eqn:sig2_sim_wdlf}
\sigma^2_{\wdlf^{k}_{\mathrm{sim}}} = \frac{\displaystyle\sum\limits_{i=1}^{N_k} \sigma^2_{n_i^k}}{(\delta\mbol^k)^2},
\end{equation}
where $n_i^k$ is the number of real stars represented by the $i^{\mathrm{th}}$ simulated
star in bolometric magnitude bin $k$.
At this point, we measure the goodness of fit between the simulated and observed WDLFs using equation \ref{eqn:chi2},
and check for convergence of the algorithm. If convergence has been reached, then $\sfr_0$ is the solution for the
star formation rate.

Next, the number of real stars that each simulated star represents is scaled so that $\wdlf_{\mathrm{sim}} = \wdlf_{\mathrm{obs}}$,
i.e. for star $i$ in bolometric magnitude bin $k$
\[
n_i \rightarrow n_i^{\prime} = n_i \times \frac{\wdlf_{\mathrm{obs}}^k}{\wdlf_{\mathrm{sim}}^k}
\]
and
\[
\sigma^2_{n_i} \rightarrow \sigma^{\prime2}_{n_i} = n_i^2 \times \sigma^2_{\frac{\wdlf_{\mathrm{obs}}^k}{\wdlf_{\mathrm{sim}}^k}} + \frac{\wdlf_{\mathrm{obs}}^k}{\wdlf_{\mathrm{sim}}^k}^2 \times \sigma^2_{n_i}
\]

This results in a simulated population of WD progenitors for which the present-day WDLF obtained using equation \ref{eqn:sim_wdlf}
exactly equals the observed WDLF, and for which the uncertainty is purely statistical arising from the finite number
of simulated stars. This is not strictly appropriate, as the error on the simulated WDLF and recovered SFR could be driven
arbitrarily low by using a large enough number of simulation stars.
In fact, we wish to assign uncertainties to our simulated stars in such a way that the
uncertainty on the simulated WDLF matches that on the observed WDLF, in the limit of a large number of simulation stars.
This is the best that could be achieved, given the errors on the data.
Any finite number of simulation stars will result in increased uncertainty, due to the additional contribution from
statistical errors.
This is achieved in a given bolometric magnitude bin $k$ by adding the term
\begin{equation}
\label{eq:errors}
\sigma^2_{n_i} \rightarrow \sigma^{\prime2}_{n_i} = \sigma^2_{n_i} + \frac{ (\delta\mbol^k)^2 \sigma^2_{\wdlf_{\mathrm{obs}}^k}} {N_k} \frac{n_i}{\langle n_i \rangle}
\end{equation}
where $N_k$ is the number of simulated stars in bin $k$, and $\langle n_i \rangle$ is the mean number of real stars that each simulated
star represents.
Inserting this term alone into the equation for the error on the simulated WDLF (equation \ref{eqn:sig2_sim_wdlf})
gives the desired result $\sigma^2_{\wdlf^{k}_{\mathrm{sim}}} = \sigma^2_{\wdlf^{k}_{\mathrm{obs}}}$ in the limit of zero statistical error.


The updated star formation rate $\sfr_1$ and formal error $\sigma^2_{\sfr_1}$ can now be obtained from the simulated star population using 
the equations
\begin{equation}
\label{eqn:sim_sfr}
\sfr_1^j = \frac{\displaystyle\sum\limits_{i=1}^{N_{\mathrm{obs}}} n_i^j }{\delta\tform}  \left(\frac{N}{N^j_{\mathrm{obs}}}\right) \left(\frac{1}{1 - A_j}\right)
\end{equation}
\begin{equation}
\label{eqn:sig2_sim_sfr}
\sigma^2_{\sfr_1^j} = \frac{\displaystyle\sum\limits_{i=1}^{N} \sigma^2_{n_i^j}}{(\delta\tform)^2} \left(\frac{N}{N^j_{\mathrm{obs}}}\right)^2  \left(\frac{1}{1 - A_j}\right)^2,
\end{equation}
where $n_i^j$ is the number of real stars represented by the $i^{\mathrm{th}}$ simulated star in formation time bin $j$.
The sum includes only observed stars, \ie stars that form white dwarfs that lie within the range of the observed WDLF at the present day.
The factor $\frac{N}{N^j_{\mathrm{obs}}}$ corrects the rate for the fraction of missing stars, where $N$ is the number of
simulation stars in each formation time bin, and $N^j_{\mathrm{obs}}$ is the number of observed simulation stars in bin $j$.
The factor $\frac{1}{1 - A_j}$ accounts for low mass stars that form in bin $j$ that don't produce white
dwarfs at the present day, and is calculated by
\begin{equation}
A_j = \frac{1}{\delta\tform} \int_{\delta\tform^j}  \int_{0.6\msolar}^{\msmassfnt} \! \imf(\mms) \, \mathrm{d} \mms   \mathrm{d}\tform
\end{equation}
This is analogous to equation \ref{eqn:low_mass_factor} in the continuous case.

%
\subsection{Input parameters}
\subsubsection{The initial mass function}
In this study, the initial mass function is a simple power law with exponent $-2.3$. This is appropriate for stars
more massive than $0.5\msolar$ \citep{kroupa2001}, which encompasses the entire range of interest.
\subsubsection{Main sequence lifetimes}
In order to provide an estimate of the main sequence lifetime as a function of progenitor mass, we adopt the 
stellar evolutionary model grid of the Padova group \citep{Bertelli2008,Bertelli2009}. These cover the mass
range $0.15 - 20 \msolar$ for 39 different metal and helium abundances.
We consider the pre-white dwarf phase to last from the zero age main sequence to the first thermal pulse,
and for low mass stars we include the time spent on the horizontal branch.

Lifetimes at arbitrary mass are interpolated between grid points using a simple linear scheme; we experimented with cubic splines and
functions of the form $\tms(M) = AM^{-2.5B}$, and found a linear approach to be a good trade off between accurately
representing the data and avoiding artefacts. The metallicity is not interpolated, and for each simulated population
we assume a single constant value 
of $Z$ = 0.017 for the metal content and $Y$ = 0.30 for the helium content.
%
%
\subsubsection{The initial-final mass relation}
The initial-final mass relation $\ifmr$ determines the mass of the white dwarf $\mwd$ that forms from a progenitor
of mass $\mms$. In this work we consider a range of IFMRs, and adopt the empirical linear model of \citet{kalirai2008}
as a baseline. This relation has the form
%
%
%
\begin{equation}
\mwd = 0.109 \mms + 0.428
\end{equation}
In conjunction with the upper mass limit for WD formation, this relation fixes the mass of the heaviest WD at $1.19M_{\odot}$.
At low masses, this relation breaks down at $\mwd=\mms\sim0.48M_{\odot}$. However, stars of this mass have ages well in excess
of the age of the universe and do not contribute to our models.
\subsubsection{White dwarf cooling times}
\label{wd_models}
White dwarf luminosities at a given cooling time, mass and atmosphere composition are obtained 
by interpolating grids of model cooling sequences.
In order to check the sensitivity of the results to the WD models, we use two independent sets.

%
%
%
%
The first set of models is that of the Montreal group. 
We use the latest set of cooling sequences available online\footnote{See \texttt{www.astro.umontreal.ca/$\sim$bergeron/CoolingModels}}
at the time of writing: these are described in \citet{fontaine2001},\citet{bergeron2001} and updated in 
\citet{Tremblay2011},\citet{Bergeron2011}
and references therein.
Note that we only require bolometric magnitude as a function of cooling time: colours in specific filter systems and synthetic spectra are
not used.
These cooling sequences will be referred to as the F01 models from now on.
%
%
The F01 models are computed using pure carbon cores at high temperatures ($T_{\mathrm{eff}}>30,000$K),
and uniformly mixed carbon/oxygen cores of equal mass fraction at lower temperatures.
The core composition is important because the rate of cooling is determined, among other things,
by the ionic specific heat.
In these models, the additional energy source at low temperatures associated with the sedimentation of carbon and
oxygen upon crystallisation of the core is not included.
%
The hydrogen atmosphere models have standard `thick' envelopes, consisting of an outer H layer
of mass fraction $q_{\mathrm{H}}=10^{-4}$ on top of a He layer of mass fraction $q_{\mathrm{He}}=10^{-2}$.
The helium atmosphere models are similar, but with $q_{\mathrm{H}}=10^{-10}$.
%
%
For both atmosphere types, models are computed at constant mass over the range 0.2--1.2$\msolar$ in steps of 0.1$\msolar$.
%
For each mass, the cooling time varies from up to several Myr to 15 (8) Gyr for H (He) models,
over which time the bolometric magnitude varies approximately 0(6)--20(17).
%
For the He models, a discontinuity exists between around $25000K$ and $35000K$ which we 
interpolate over. This is due to the use of two different sets of evolutionary models at high and
low temperatures, which match well on stellar radii but have small discontinuities in the cooling times
at the point where they are joined
(Bergeron, priv. comm.). No such discontinuity is present in the H models.
%
%
%
%
%
%

The second set of models is taken from the BaSTI database\footnote{Official website at \texttt{http://www.oa-teramo.inaf.it/BASTI}}
and is described in \citet{Salaris2010}.
%
For these models, the core is composed of a carbon and oxygen mixture with relative abundance and distribution
that varies as a function of WD mass (see Salaris et al fig. 2). For each WD mass, the C/O stratification 
is taken from a connecting main sequence evolutionary model at the first thermal pulse (obtained
from the BaSTI database for Solar metallicity).
%
Cooling sequences are computed with and without the effects of C/O phase separation and sedimentation on crystallisation, which 
slows the cooling of stars at low luminosities. These models will be referred to as S10 and S10p (including C/O phase separation effects)
from now on.
%
For the hydrogen atmosphere models, the envelope consists of a surface H layer with mass 
fraction $q_{\mathrm{H}}=10^{-4}$ on top of a $q_{\mathrm{He}}=10^{-2}$ He layer (the same as for F01),
and the helium atmosphere models have an envelope consisting of a single $q_{\mathrm{He}}=10^{-3.5}$ He layer.
%
%
%
Nine discrete masses are computed in the range 0.54--1.2$\msolar$, in intervals varying
0.01--1.0 $\msolar$. Models are denser in the 0.5--0.8$\msolar$ range, to reflect the general
higher abundance of these stars.
%
%
For each mass, cooling time varies from around several Myr to 15 (8) Gyr for H (He) models, and
the bolometric magnitude varies approximately 3--16 for both types.
%
%
%
%
%
Two representative cooling tracks are shown in Fig. \ref{fig:WDs} comparing the F01, S10 and S10p models.
\begin{figure}
\centering
\includegraphics[trim=0cm 0cm 0cm 0, clip=true, width=8.5cm]{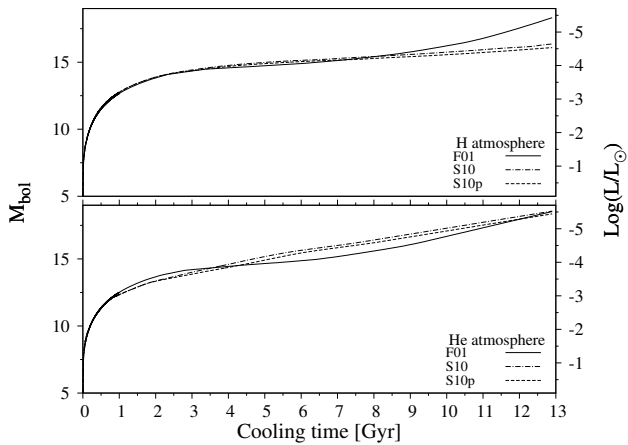}
\caption[]{Cooling tracks for 0.6$\msolar$ WDs, for each set of evolutionary models. The solid lines show the F01 models.
The dashed lines show the S10p models (including phase separation effects); the dot-dashed line shows the S10 models
(excluding phase separation effects). Phase separation effects become significant at cooling times greater than roughly 2 Gyr.}
\label{fig:WDs}
\end{figure}

We experimented with a variety of interpolating functions, and found a bilinear scheme to be the most 
appropriate. It is sometimes necessary to extrapolate bolometric magnitudes at points outside the 
model grid, either at very early or late cooling times, or at very low masses in the case of the
S10 models. We use the same bilinear method for this, and are careful to check that
results do not rely too heavily on extrapolated points far outside the model grid.
\section{Validation with Synthetic Data}
\subsection{Synthetic Data Generation}
In order to test the performance of the inversion algorithm, we run it on a set of synthetic WDLFs
derived from known input star formation histories. Synthetic WDLFs for test purposes can be calculated
in the same way as is done during the inversion algorithm, \ie generating a synthetic population of WDs and 
using equations \ref{eqn:sim_wdlf} and \ref{eqn:sig2_sim_wdlf}.
A WDLF generated by following the algorithm up to that point is strictly appropriate for a \textit{volume limited}
sample of WDs; each WD that is present is used to determine the WDLF without selection according to magnitude.
This is the best method for the inversion algorithm, when good constraint over the entire WDLF range is required.
However, this isn't realistic for modelling observational WDLFs: these are typically derived from \textit{magnitude limited}
catalogues and have quite different noise profiles. The main difference is that the statistical noise at the faint end
is much greater and the WDLF generally doesn't extend as far, due to intrinsically faint stars being preferentially lost.
Figure \ref{fig:synWDLFs} shows two synthetic WDLFs derived from volume and magnitude limited WD populations.
It is important that our sensitivity and validation
tests consider realistic noise models in order to get a true estimate of the performance of the algorithm on real data.

We simulate a magnitude limited WDLF in the following manner. First, we select an appropriate apparent magnitude limit and use the
minimum absolute magnitude of any simulated WD to determine the theoretical survey edge. In conjunction with a model
of the density profile, this defines the total volume of the survey region $V_{\mathrm{tot}}$.
Then, for each simulated WD we calculate the maximum observable distance, and use this to determine the accessible survey
volume $V_{\mathrm{max}}$ in which the star is observable. This is used to assign an observation probability
\[
p_{\mathrm{obs}} = \frac{V_{\mathrm{max}}}{V_{\mathrm{tot}}},
\]
assuming stars are distributed uniformly within the survey volume. We use $p_{\mathrm{obs}}$ to randomly
select or reject each simulated WD; those that are selected have their density contribution scaled according to
\[
n_i \rightarrow n_i^{\prime} = n_i \times p_{\mathrm{obs}}^{-1}
\]
and
\[
\sigma^2_{n_i} \rightarrow \sigma^{\prime2}_{n_i} = \sigma^2_{n_i} \times  p_{\mathrm{obs}}^{-2}
\]
to correct for the missing stars. This is a variation on the standard $V_{\mathrm{max}}^{-1}$ technique that 
ensures the WDLF spatial density is dimensionless, as required.
\begin{figure}
\centering
\includegraphics[trim=1.5cm 0cm 1.8cm 0, clip=true, width=8cm]{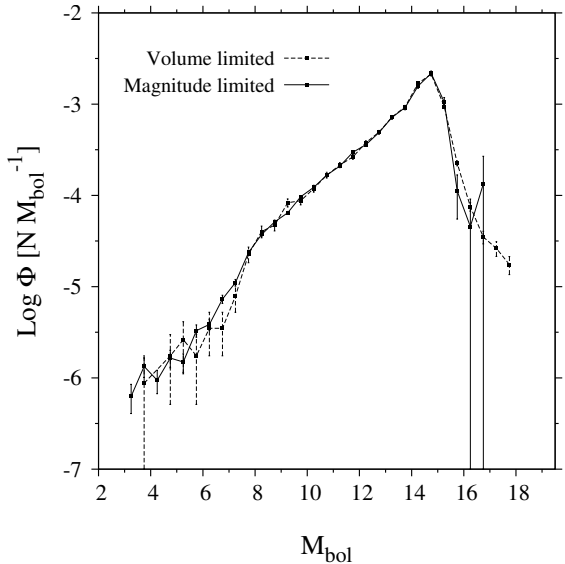}
\caption[]{Synthetic WDLFs derived for volume and magnitude limited samples of $10^4$ WDs, for a constant star formation rate. Relative to
a volume limited sample, magnitude limited samples show lower errors at the bright end and much larger errors at the faint end, due to
magnitude selection effects leading to these stars being over- and under-represented, respectively.}
\label{fig:synWDLFs}
\end{figure}
\subsection{Convergence tests}
\label{convergence}
%
%
%
%
The first round of tests is designed to check the convergence of the algorithm under a range of different star formation
rate scenarios. The particular set of scenarios chosen is summarised in Table \ref{tab:sfr}; these are intended
to cover a wide variety of stellar population types.
In all cases, the initial guess star formation rate model is a flat rate with a maximum lookback time of $13$ Gyr,
divided into 100 bins of approximately $130$ Myr width\footnote{The minimum lookback time of the initial guess rate
is slightly greater than zero ($\sim10^{7-8}$ yr), due to the finite lifetime of the most massive WD progenitors, so
bins are slightly narrower than $130$ Myr.}.
\begin{table}
\begin{center}
\caption[Total Integrated Star Formation]{Star Formation Rate models considered.}
\begin{tabular}{ll}
\hline\label{tab:sfr}
Type & Integrated rate [stars] \\
\hline
\hline
Constant          & $9\times10^{-3}$ \\
Exponential decay & $9.89\times10^{-3}$ \\
Single burst      & $1.5\times10^{-3}$ \\
Fractal		  & $5.039\times10^{-2}$ \\
\hline
\end{tabular}
\end{center}
\end{table}
These tests use synthetic WDLFs generated from a large ($2\times10^6$) WD sample with no bolometric magnitude
error. WDLFs are computed at a resolution and magnitude range that matches recent determinations of the WDLF in the Solar neighbourhood
($\mbol=1$ to $\mbol=18$, $\Delta\mbol=0.5$, \eg \citet{rowell2011}), but are otherwise noise-free, and the inversion algorithm 
uses the same set of input parameters as those used to generate the WDLFs. These are listed in Table \ref{tab:params}.
\begin{table}
\begin{center}
\caption[Modelling parameters]{Input physics used in convergence tests.}
\begin{tabular}{ll}
\hline\label{tab:params}
Parameter & Value \\
\hline
\hline
IMF exponent     & -2.3  \\
Metallicity      &       \\
......Z          & 0.017 \\
......Y	         & 0.30  \\
Initial-Final    & $\mwd = 0.109 \mms + 0.428$ \\
\hspace{0.1cm}Mass Relation    & \hspace{1em}\citep{kalirai2008} \\
$\alpha$	 & 0.8    \\
WD cooling sequences & F01    \\
\hline
\end{tabular}
\end{center}
\end{table}
%

%
%
%
The results of the inversion are shown in Fig. \ref{fig:convergence}. In each case, the first three iterations are plotted along with the
final converged fit, for both the SFR model and the WDLF. The algorithm achieves a reasonably good approximation to the underlying
SFR in all cases, though with varying success. In all cases, the recent ($\tform<5$ Gyr) SFR is accurately recovered. The constant
SFR model (\ref{fig:converged_constant}) shows a significant deviation at earlier times, and the onset of star formation
is not well resolved. However, the total integrated SFR of $8.97 \times 10^{-3}$ is within $0.4\%$ of the true value.
The exponentially decaying model is similar, and has an integrated SFR ($9.89 \times 10^{-3}$) that is within $0.1\%$ of the true value.
In the case of the single burst model, the shape of the burst is poorly resolved but the location of the peak is well recovered,
and the integrated rate ($1.53 \times 10^{-3}$) is within $2\%$ of the true value.
The fractal SFR fit reveals some important behaviour of the algorithm. High frequency components and discontinuities in 
the SFR are only well resolved at recent times; after around $4$ Gyr any features narrower than several Gyr are lost, and the 
converged fit approximates a moving average of the time varying SFR. The integrated SFR of $5.007 \times 10^{-2}$ is 
within $0.7\%$ of the true value.
In all cases, the onset of star formation is resolved to around $\sim1$ Gyr.

\begin{figure*}
\begin{minipage}{160mm}
\centering
\subfigure[Constant SFR]{
\includegraphics[height=4.75cm]{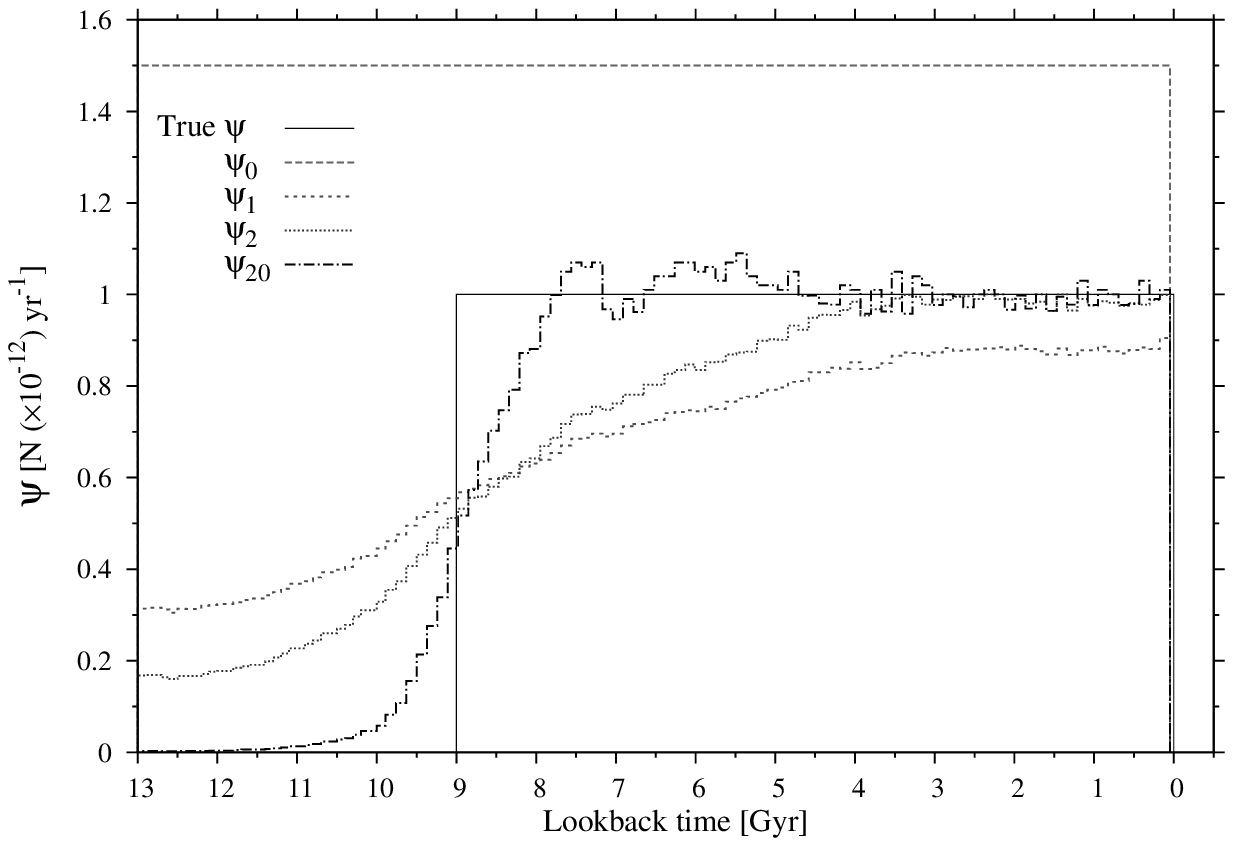}
\includegraphics[height=4.75cm]{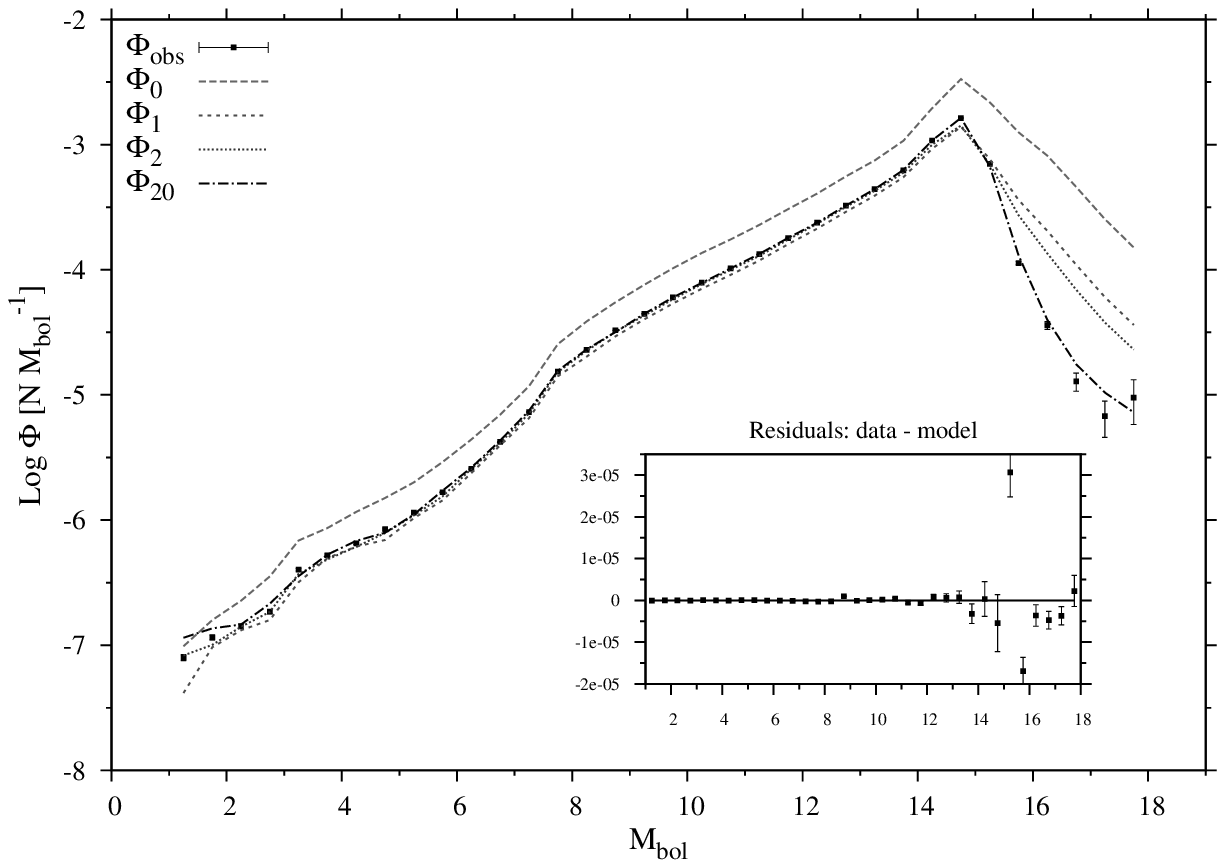}\label{fig:converged_constant}}\\
\subfigure[Exponentially decaying SFR]{
\includegraphics[height=4.75cm]{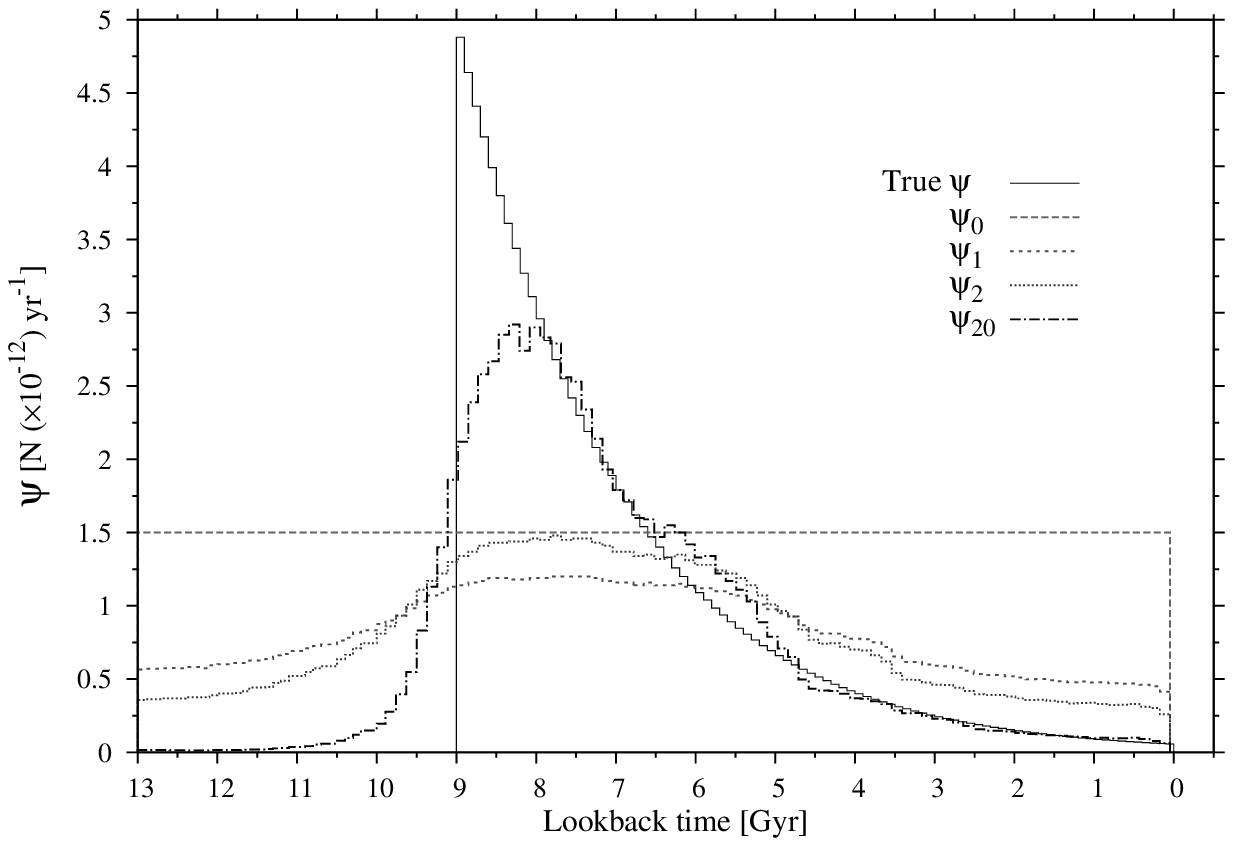}
\includegraphics[height=4.75cm]{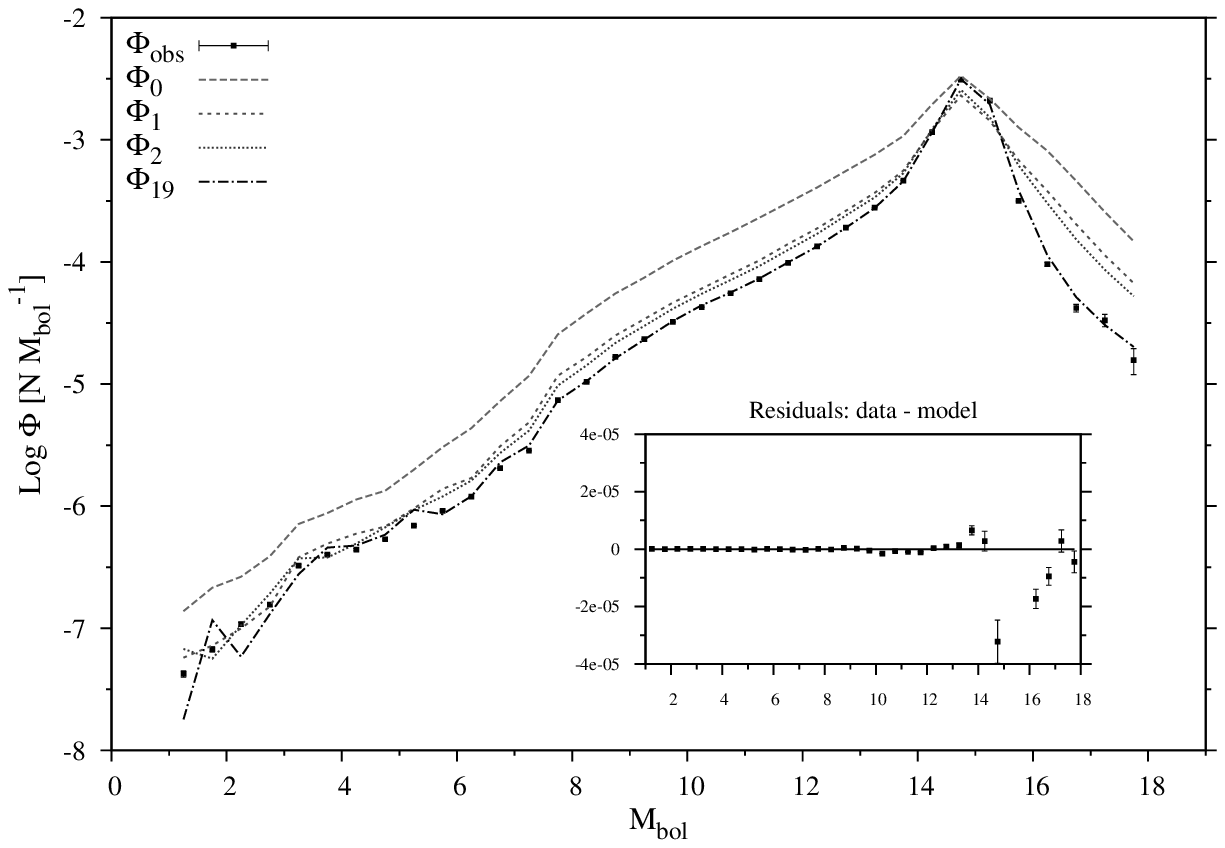}}\\
\subfigure[Single burst SFR]{
\includegraphics[height=4.75cm]{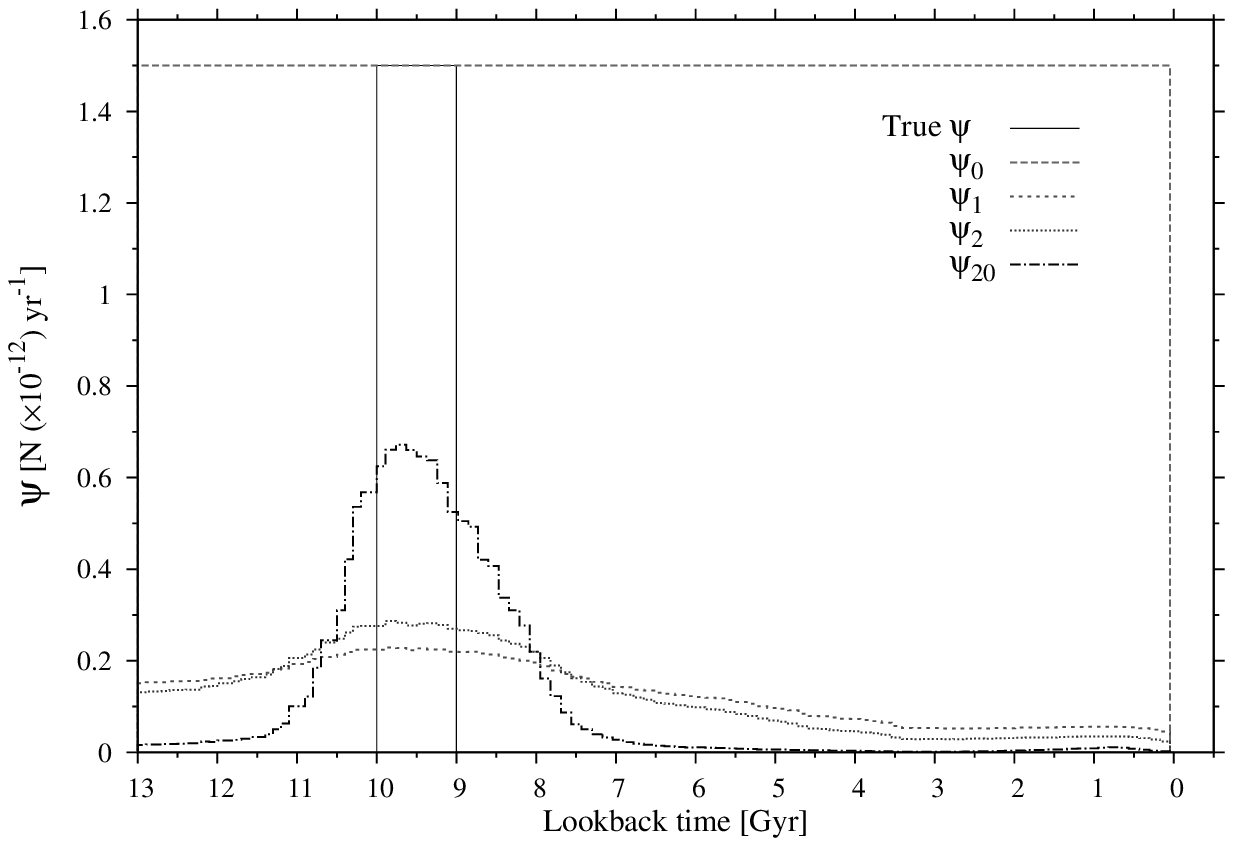}
\includegraphics[height=4.75cm]{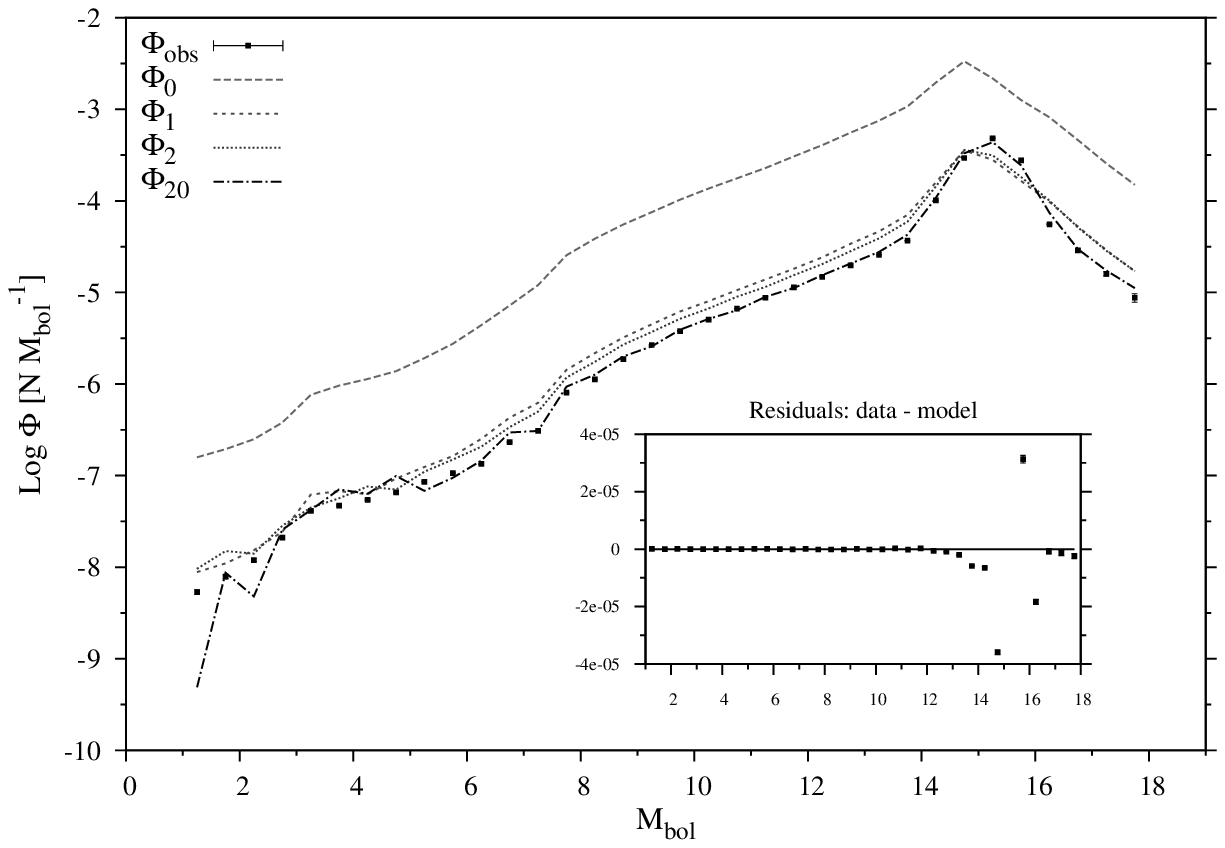}}\\
\subfigure[Fractal SFR]{
\includegraphics[height=4.75cm]{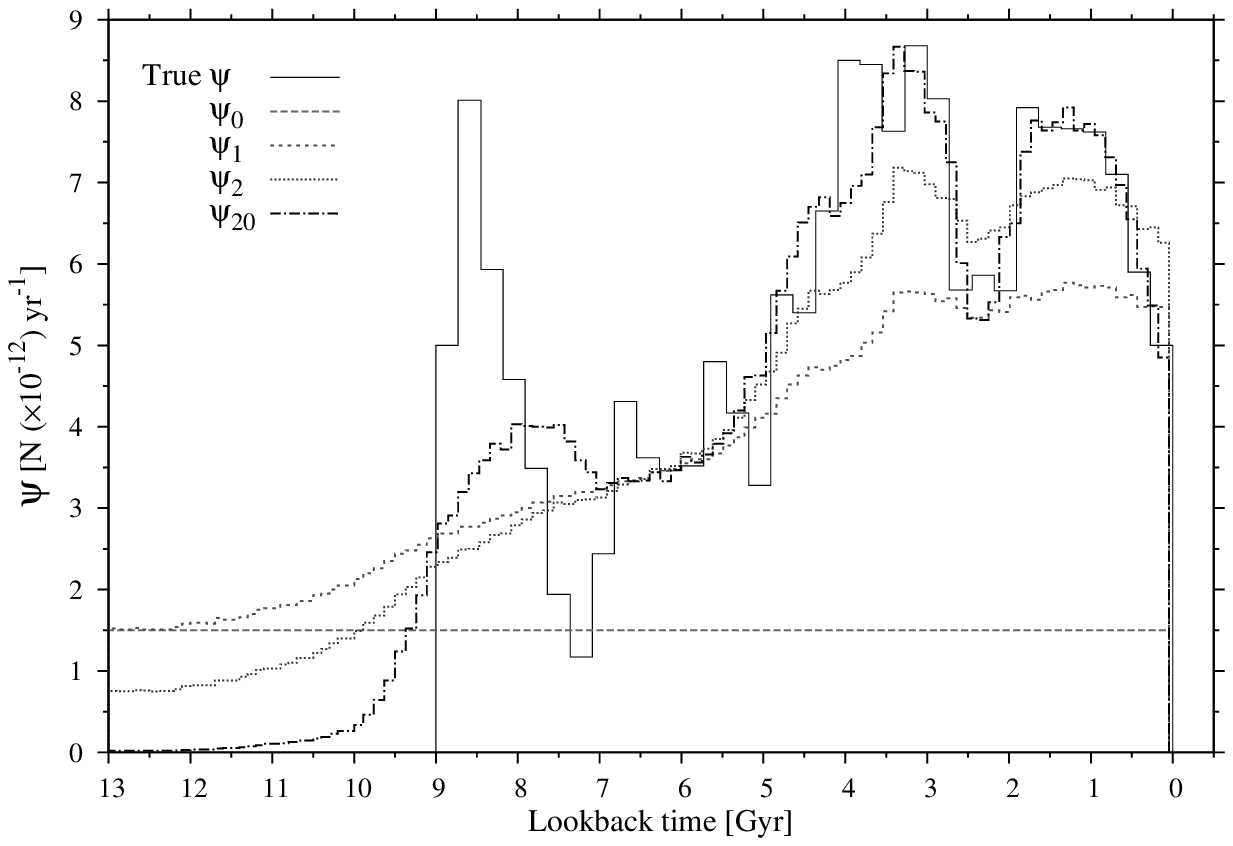}
\includegraphics[height=4.75cm]{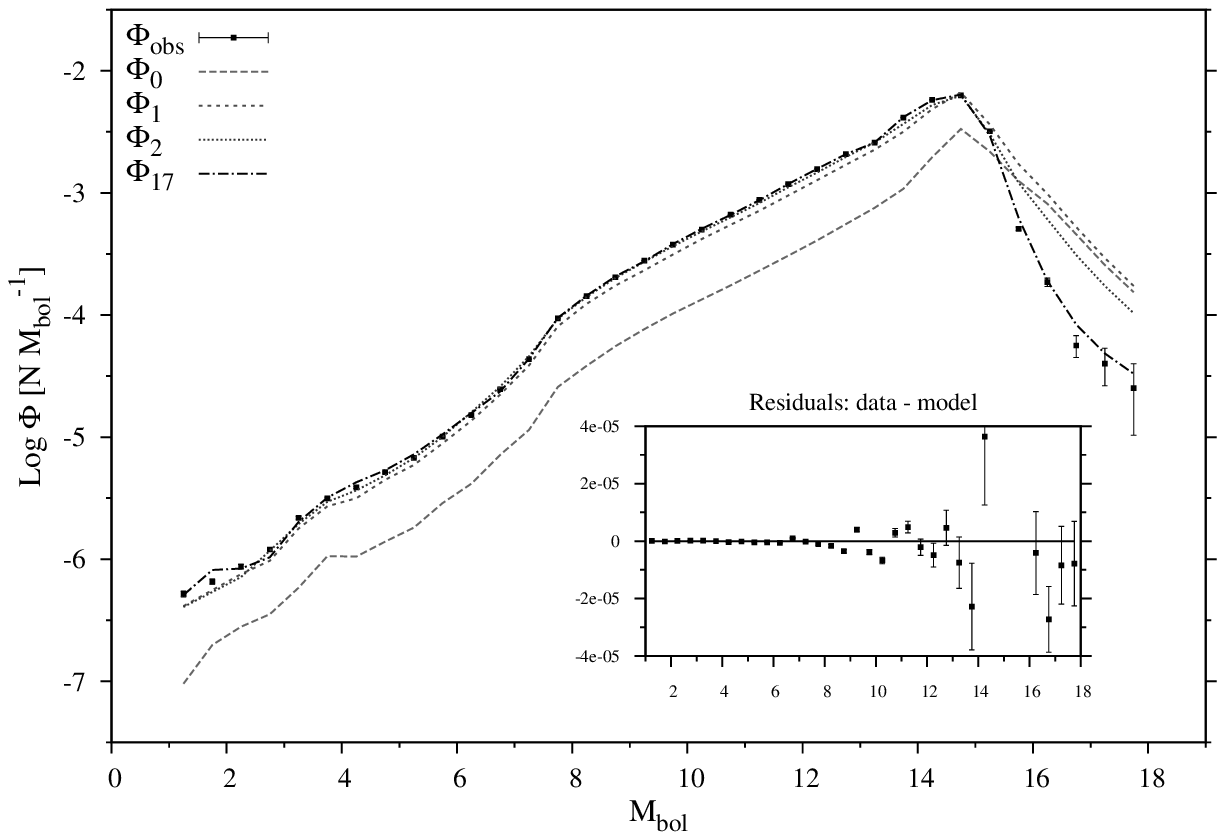}}
\caption[Convergence Tests]{Results of algorithm convergence tests on noise-free data. In each case, the panel on the left shows the
recovered SFR model for the first 3 iterations (dashed grey) and the converged solution (dot-dashed), 
along with the ground truth (solid). The panel on the
right shows the simulated WDLF derived for the same iterations, along with the observed WDLF used to constrain the algorithm.
The inset panel shows the residuals for the final converged WDLF.}
\label{fig:convergence}
\end{minipage}
\end{figure*}

We investigated the effect on the SFR solution of using fainter ($\mbol \rightarrow 24$) and higher resolution ($\Delta\mbol=0.25$) WDLFs, and
of setting the maximum lookback time equal to the time since the onset of star formation (for the case of measuring the time variation
in star formation for a population of known total age). For both the faint and the high resolution WDLFs, no significant difference is seen
around the onset of star formation, although the fainter WDLF resolves the total age slightly better. The artefact in the constant and
exponentially decaying SFR solutions is still present. The narrow WDLF provides better overall SFR resolution, and in the
case of the fractal SFR more of the fine detail is recovered at all ages. Setting the maximum lookback time equal to the total
age of the population eradicates the artefact in the constant and exponentially decaying SFR solutions, and results in considerably better
constraint at older times in all SFR models. However, in the case of the single burst model the falling edge of the burst is still not well resolved.

To summarize, the algorithm is effective at measuring the average time varying SFR for a range of population types, with greater
resolution at more recent times depending mainly on the magnitude resolution of the WDLF. The total integrated SFR is correctly recovered.
The algorithm does not handle discontinuities well,
such as sudden bursts or lulls in star formation, though if the total population age is known this can be used to avoid artefacts
around the onset of star formation.
\subsection{Noise Degradation}
\label{noise}
Inverse problems are notoriously susceptible to noise. In the case of the WDLF, this will manifest as errors
on both the number density at a given magnitude, and the bolometric magnitude of individual stars, resulting
in an overall smoothing. It is important to estimate the effect that these errors have
on the inversion procedure, given that the observed WDLF will be subject to both types to some degree.

We have repeated the tests of section \ref{convergence}, using simulated WDLFs with realistic error models.
These were calculated by reducing the number of simulated WDs to 10$^4$, to give a statistical uncertainty
on the number density that matches the WDLFs of \cite{harris2006} and \cite{rowell2011}. Bolometric magnitude
errors were added to each star by drawing from a Gaussian distribution of width 0.25. This is the approximate
size of errors in the \cite{rowell2011} WDLF; the \cite{harris2006} errors are likely to
be smaller given the better photometry of the SDSS relative to SuperCOSMOS and the sharper features of their WDLF.

\begin{figure*}
\begin{minipage}{160mm}
\centering
\subfigure[Constant SFR]{
\includegraphics[trim=0cm 0cm 0cm 0, clip=true, height=5cm]{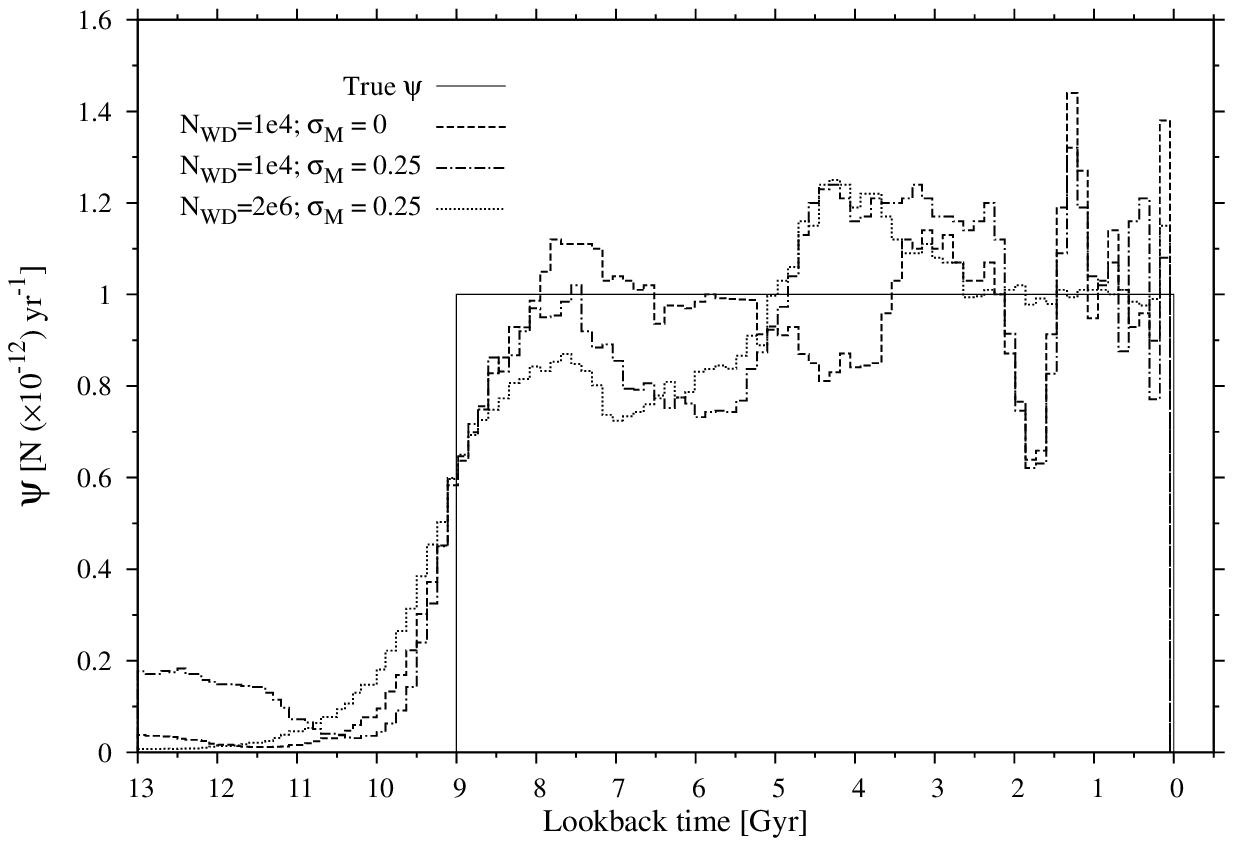}}
\subfigure[Exponentially decaying SFR]{
\includegraphics[trim=0cm 0cm 0cm 0, clip=true, height=5cm]{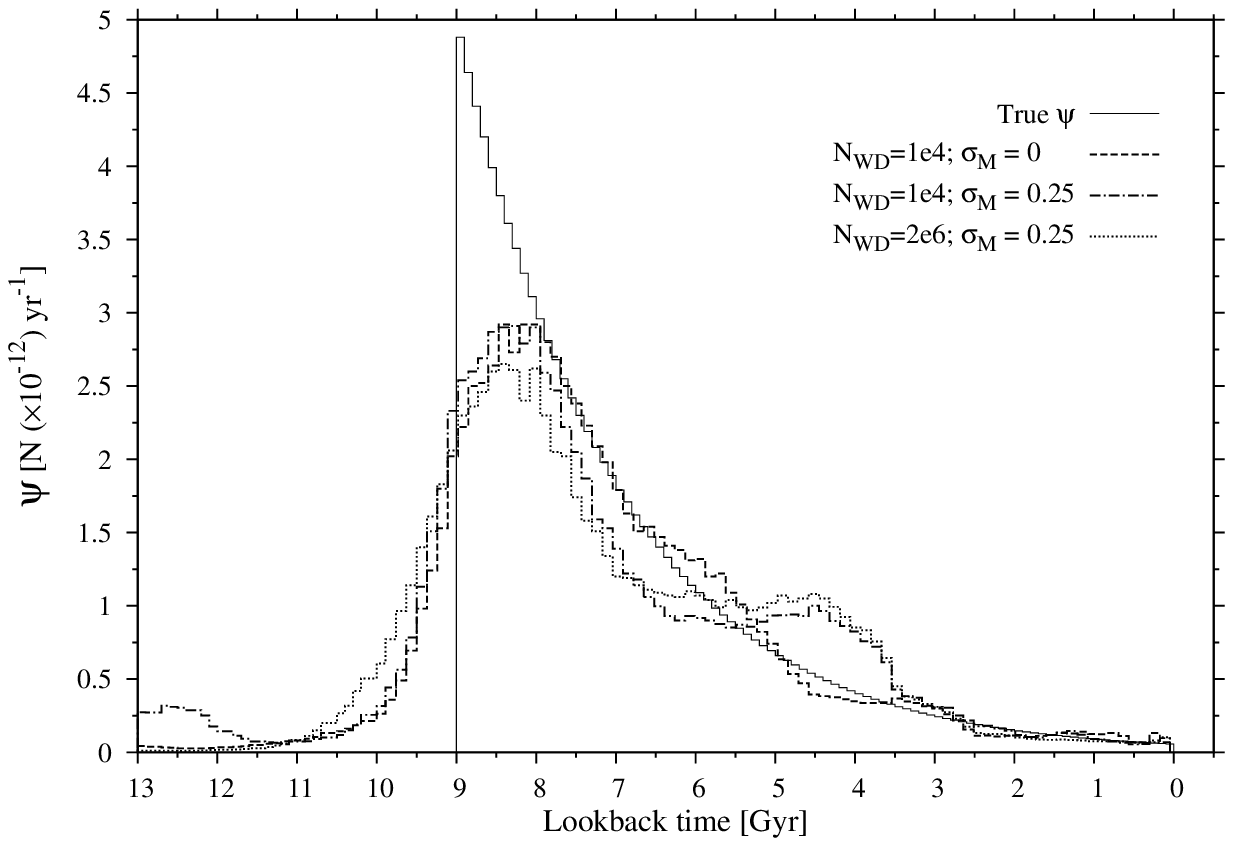}}\\
\subfigure[Single burst SFR]{
\includegraphics[trim=0cm 0cm 0cm 0, clip=true, height=5cm]{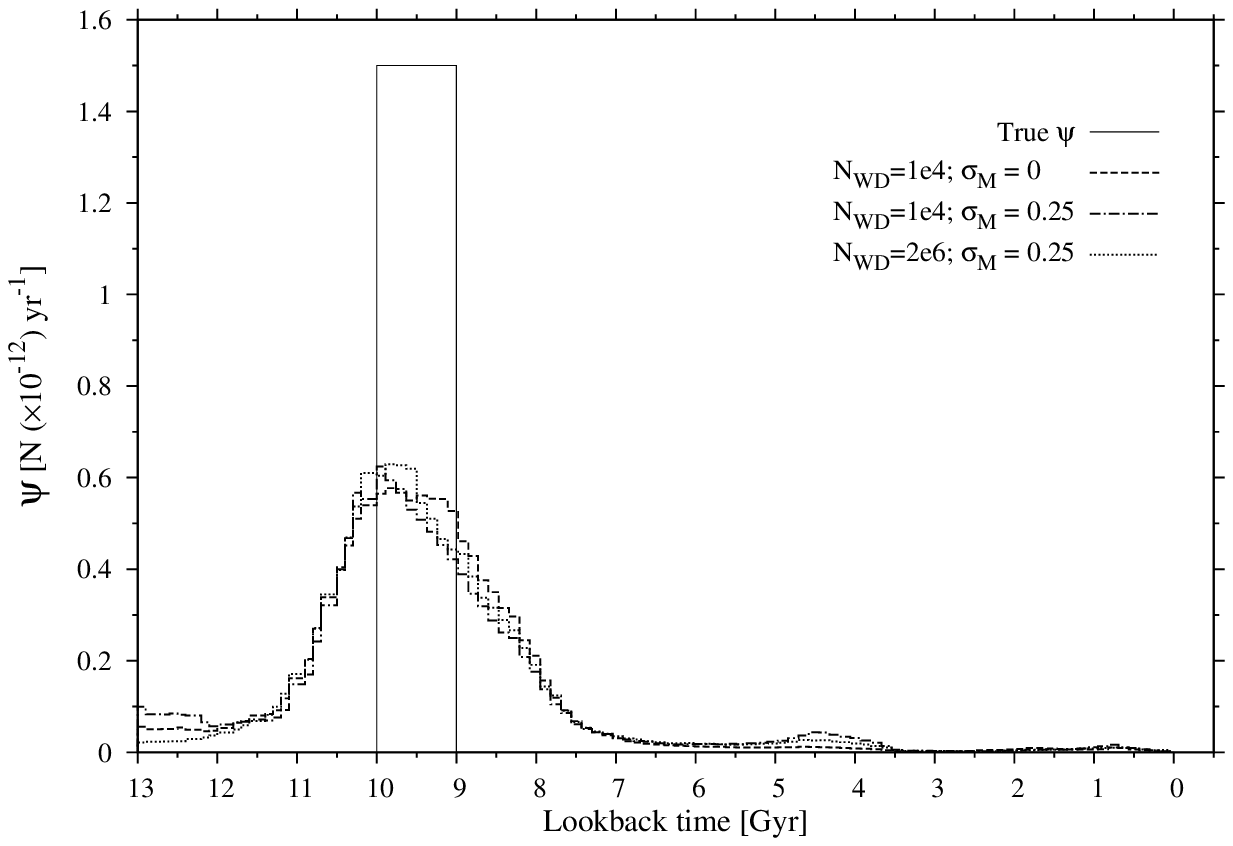}}
\subfigure[Fractal SFR]{
\includegraphics[trim=0cm 0cm 0cm 0, clip=true, height=5cm]{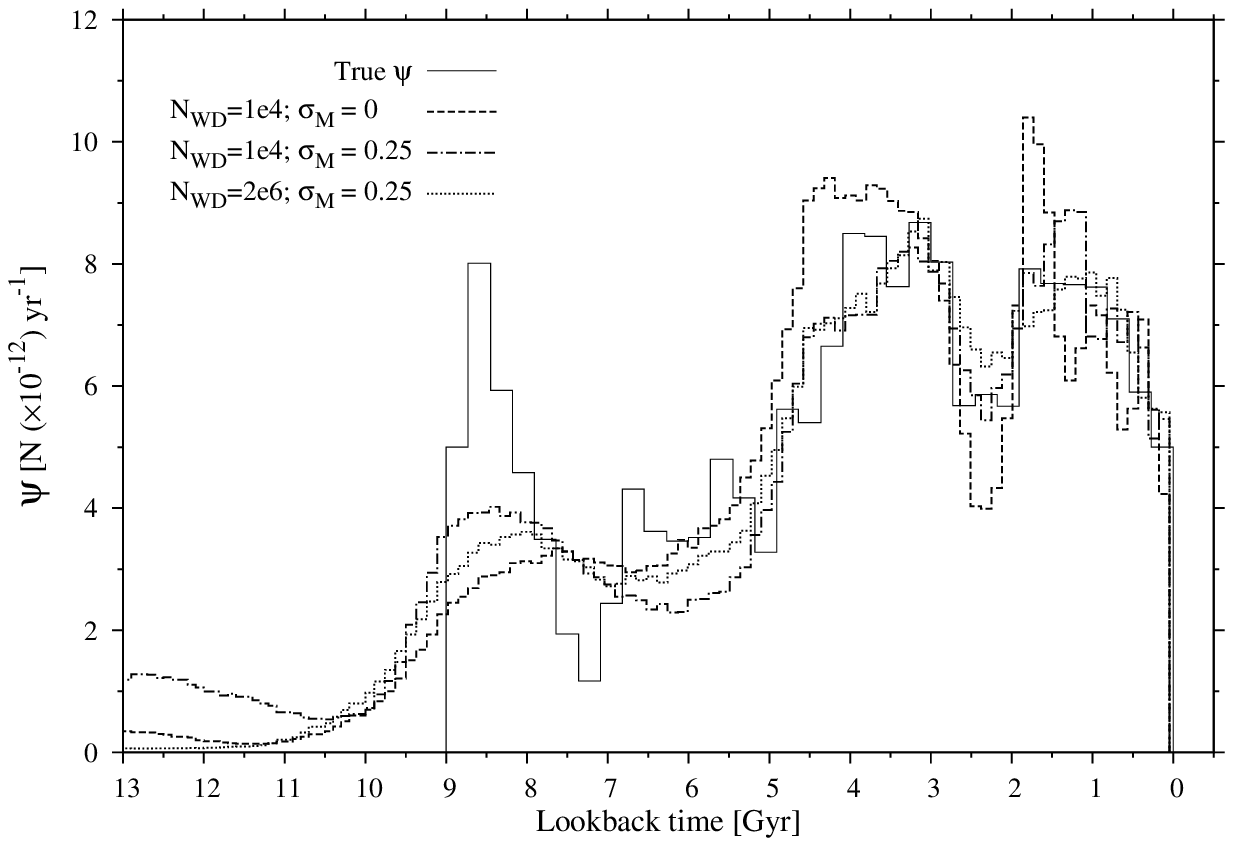}\label{error_test}}
\caption[Noise Degradation Tests]{Results of noise degradation tests on the inversion algorithm. In each case, the solid line
indicates the true underlying star formation rate. The dashed line shows the results of the inversion algorithm when applied
to a WDLF determined using $10^4$ simulated WDs with no bolometric magnitude errors. This results in statistical uncertainty on the WD density
of the same order as that of recent Solar neighbourhood WDLF measurements. The dotted line shows the results when a large
number of simulation stars (2$\times$10$^6$) are used resulting in very low density uncertainty, and bolometric magnitude errors drawn from a 
$\sigma_M = 0.25$ Gaussian distribution. The dot-dash line shows the results when both types of error are included.}
\label{fig:noise_degradation}
\end{minipage}
\end{figure*}

The effect of each type of error is shown in Fig. \ref{fig:noise_degradation}.
%
%
The presence of realistic errors in the WDLF density (green lines)
does not have a significant effect on the inversion results for both the exponentially decaying and single burst SFR cases. The constant SFR
case is quite noisy at recent times, and the fit is marginally improved in the fractal SFR case. In all cases, the integrated SFR is within
4\% of the true value.

When errors are introduced on the bolometric magnitude of stars in the observed WDLF (blue lines), the inversion results are not
significantly affected. In the case of the fractal SFR, the additional smoothing of the observed WDLF that results causes a loss
in resolution at recent times and high frequency components in the SFR are not as well recovered. Again, the integrated SFR is
within 3\% of the true value in all cases.

Errors in both the WDLF number density and bolometric magnitude of individual stars (red lines) causes additional degradation in
all cases. This is particularly bad for the constant SFR model, for which the noise at recent times is quite severe. However, the
overall form of the SFR is still reasonably well recovered, to the extent that significantly different SFR scenarios can still be
distinguished, and it is encouraging that the fractal SFR is still well recovered, albeit with rather lower resolution on the SFR
features. The integrated SFR is within 4\% of the true value in each cases.

%
\subsubsection{Error estimation}
The inversion algorithm is capable of estimating the error on the converged SFR solution, by considering the propagation of number
density errors from the observed WDLF to the corresponding SFR bins according to equation \ref{eq:errors}.
Figure \ref{fig:errors} shows again the SFR solution from
Fig. \ref{error_test} for the case of both number density and bolometric magnitude errors, and includes the estimated uncertainty
on the solution drawn in light grey. We find that the errors are generally well estimated in regions free of discontinuities, however
in the vicinity of discontinuities significant departures are present, arising from degeneracies in the inversion solution that
are not accounted for by the error estimation process.
\begin{figure}
\centering
\includegraphics[trim=1.7cm 0cm 1.7cm 0, clip=true, width=8cm]{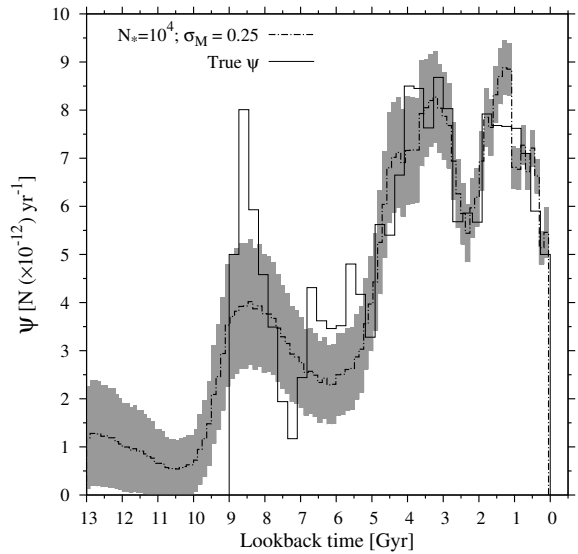}
\caption[]{Estimated error on the SFR solution of Fig. \ref{error_test}, for the case of both density and bolometric magnitude errors.
The solid line shows the true underlying SFR; the dot-dashed line and grey region shows the converged solution and estimated error. While the
uncertainty is generally well estimated, there are significant departures arising from degeneracies in the inversion. In all cases, 
these are worse at more recent times and around discontinuities.}
\label{fig:errors}
\end{figure}
%


%
\subsection{Critical Parameters}
\label{sec:crit_param}
%
%
%
%
In order to fully characterise the inversion algorithm, and to properly interpret results derived from real data, it is important to 
estimate the sensitivity of the inversion process to variations in the input parameters. Some parameters, such as the slope of the IMF
over the relevant mass range, are relatively well constrained; others, such as the fractions of hydrogen and helium atmosphere WDs and their
cooling rates, are less well constrained, and could potentially lead to systematic errors in the SFR solution.

We have repeated the convergence and noise sensitivity tests of Sections \ref{convergence} and \ref{noise}, using the same
set of synthetic WDLFs that were calculated using the parameters listed in Table \ref{tab:params}. In the present tests, we vary the parameters that
the inversion algorithm uses, and compare the results using the incorrect parameters with those obtained using the true values.
This involves a large number of individual tests, from which we present a selection of results in Fig. \ref{fig:critical_parameters}.
These were obtained for the fractal SFR model, using a synthetic WDLF that includes both density and bolometric magnitude errors.
The results are similar for the remaining SFR models, and for the noise-free WDLFs.

\begin{figure*}
\begin{minipage}{160mm}
\centering
\subfigure[$\alpha$ (ratio H/(H+He))]{\includegraphics[trim=1.9cm 0cm 2cm 0, clip=true, height=5.3cm]{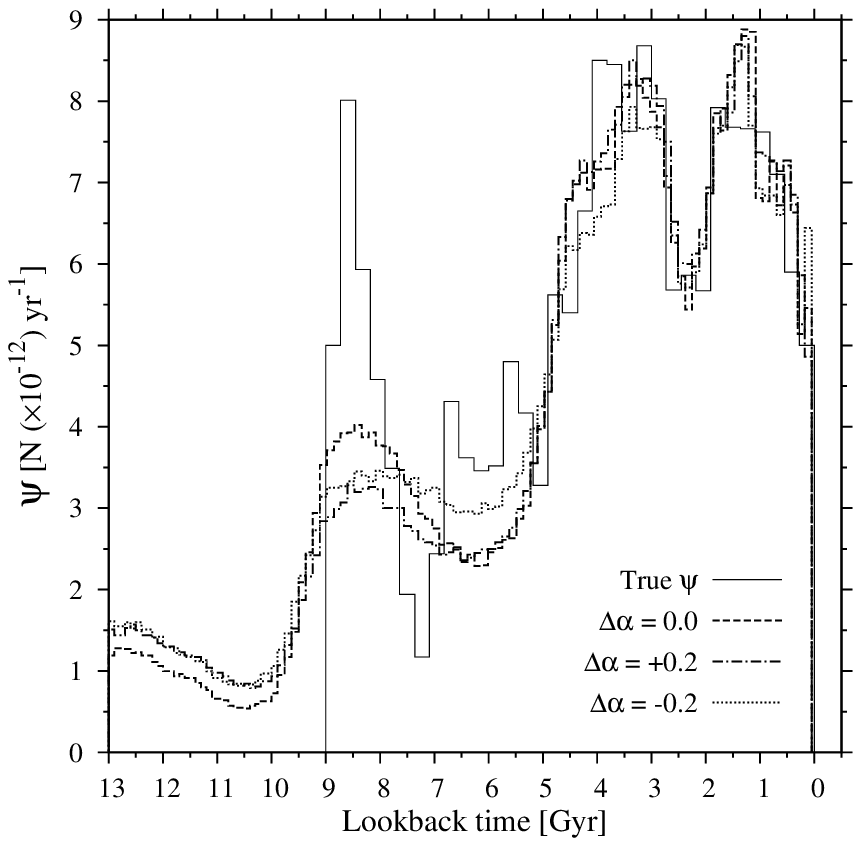}\label{fig:alpha}}
\subfigure[IFMR]{\includegraphics[trim=1.9cm 0cm 2cm 0, clip=true, height=5.3cm]{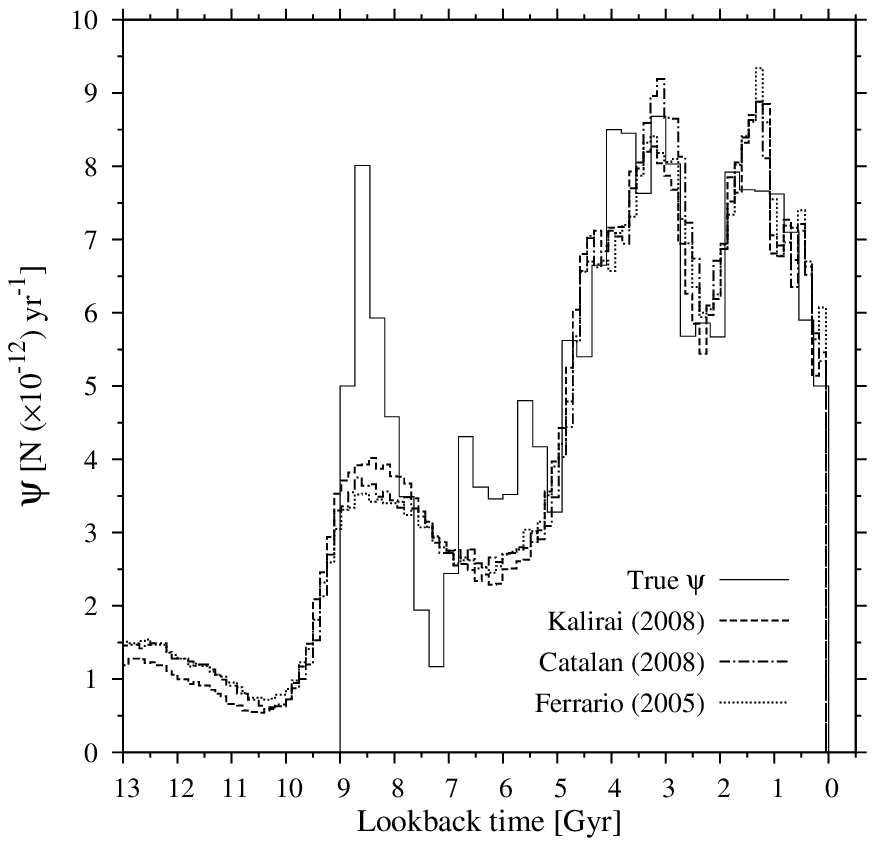}\label{fig:ifmr}}
\subfigure[IMF exponent]{\includegraphics[trim=1.9cm 0cm 2cm 0, clip=true, height=5.3cm]{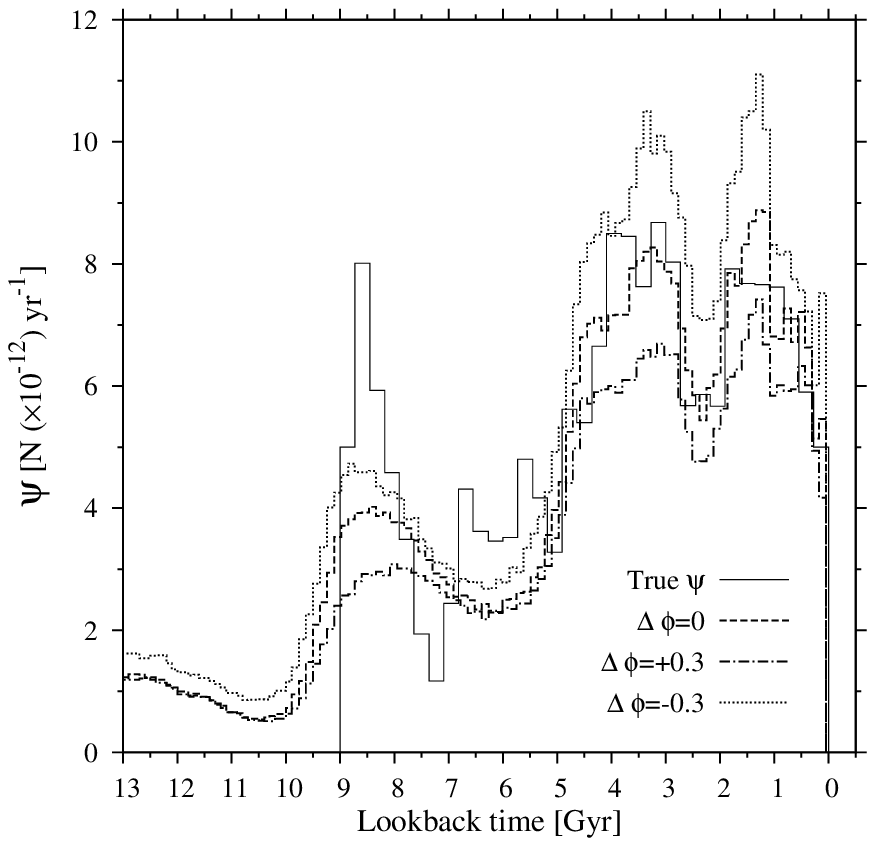}\label{fig:imf}}\\
\subfigure[Metallicity Z]{\includegraphics[trim=1.9cm 0cm 2cm 0, clip=true, height=5.3cm]{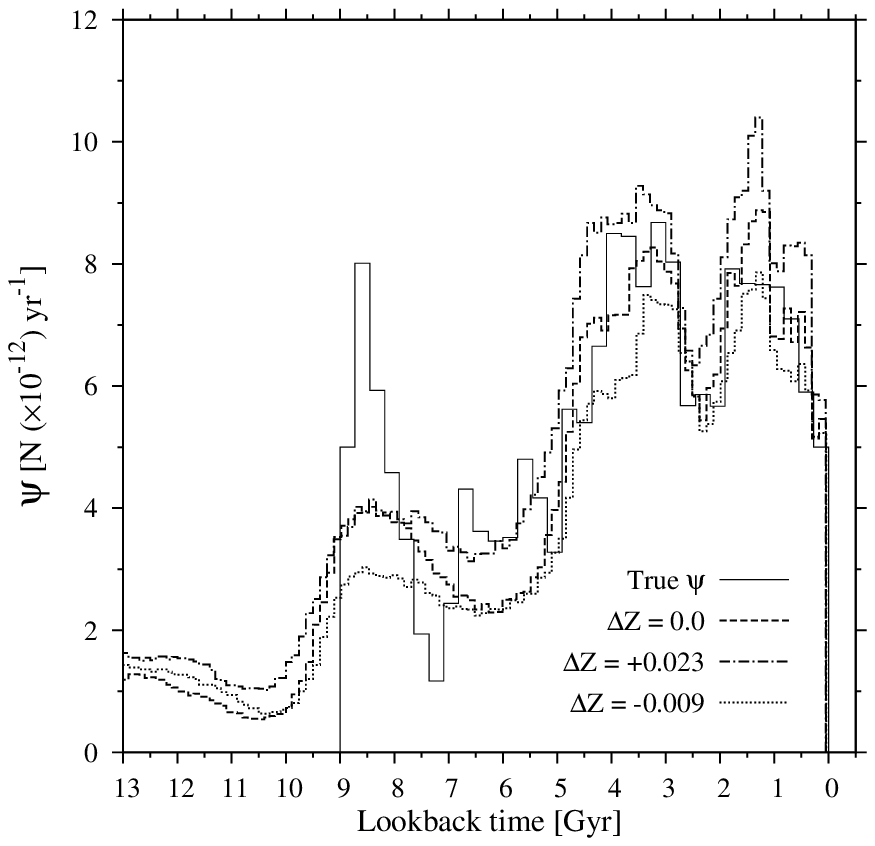}\label{fig:z}}
\subfigure[WD cooling models]{\includegraphics[trim=1.9cm 0cm 2cm 0, clip=true, height=5.3cm]{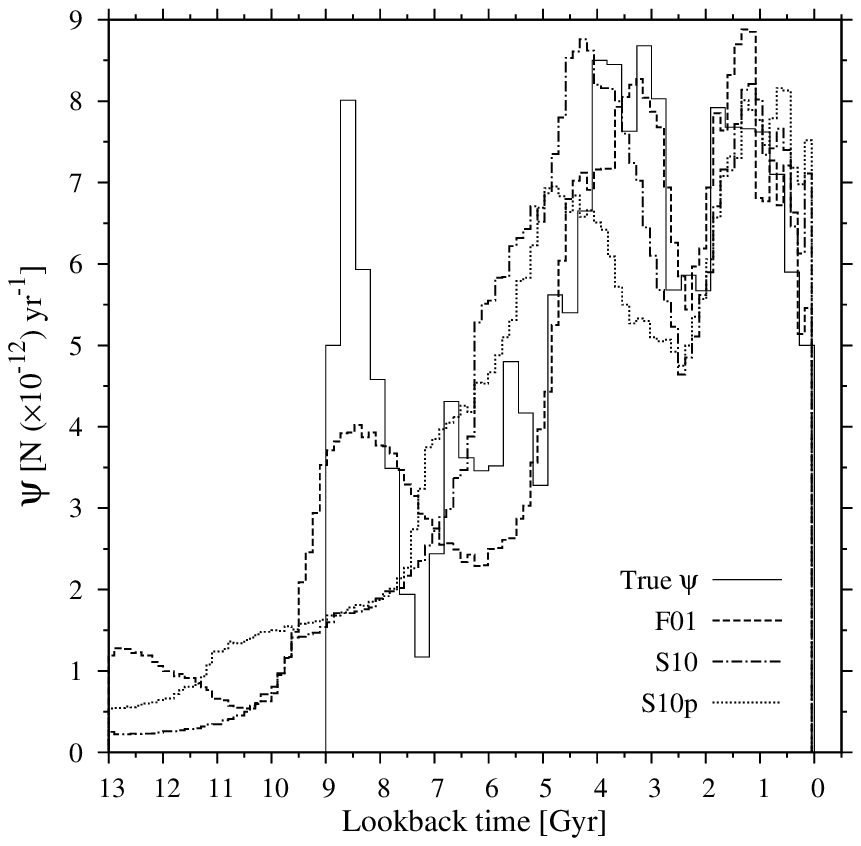}\label{fig:cooling}}
\caption[Critical Parameter Tests]{Selected results from critical parameters tests. In all cases, the solid line indicates the true underlying
SFR, and the dashed line indicates the results of the inversion algorithm when the same set of parameters is used both to generate the
synthetic WDLF, and to invert it. The dotted and dot-dashed lines show the inversion results when perturbed parameter values are used.
In these tests, a noisy synthetic WDLF is used. The algorithm is relatively insensitive to differences in the 
fraction of H atmosphere WDs within the last few Gyrs (\ref{fig:alpha}), although artefacts appear at older times due to the different cooling 
rates of the two types. Variations in the IFMR do not
have a significant effect on the performance (\ref{fig:ifmr}). Uncertainty in the IMF exponent has a significant effect on the normalisation of the SFR solution but not on the 
functional form (\ref{fig:imf}), and the effect of variations in the metallicity is very similar (\ref{fig:z}). In both cases, overproduction of
massive stars in the inversion algorithm results in an overall reduction in the SFR in order to match the observed WDLF. By far the largest
source of error in the recovered SFR arises from the choice of WD cooling models (\ref{fig:cooling}).}
\label{fig:critical_parameters}
\end{minipage}
\end{figure*}
%
%
%

%
\subsubsection{Parameter $\alpha$}
%
%
%
%
%
%
%
%
Figure \ref{fig:alpha} shows the effect of varying the fraction of hydrogen atmosphere WDs by $\pm$20\%. There is very little difference within
the last 3 Gyr, due to the cooling rates of the two types being very similar at intermediate temperatures.
At older times, differential cooling starts to become significant and the recovered SFRs show systematic deviations of up
to around 25\%. However, the integrated SFR is mostly unaffected and only deviates by less than one percent in these tests.
\subsubsection{Initial-Final Mass Relation}
%
%
%
%
%
%
%
Figure \ref{fig:ifmr} shows the effect of varying the initial-final mass relation. In these tests, the true IFMR is that of
\citet{kalirai2008}, which is derived from two old open clusters to better constrain the low mass end of the IFMR.
The first alternative IFMR is the linear fit of \citet{ferrario2005} that 
is derived from a selection of young open clusters in the Solar neighbourhood, and uses the F01 models to obtain WD masses and cooling times.
This function is shallower than the \citet{kalirai2008} IFMR.
The second alternative IFMR is that of \citet{catalan2008}, which is derived from a sample of local open clusters and common proper motion pairs
that largely overlaps with the \citet{ferrario2005} study. However, they use the white dwarf cooling models of \citet{salaris2000}
(a predecessor of the S10 models), so the analysis is largely independent. Their piecewise linear fit is steeper than the \citet{kalirai2008}
IFMR for masses greater than $2.7\msolar$, and shallower at lower masses.

It is evident from Fig. \ref{fig:ifmr} that variations in the IFMR on this level do not have a large effect on the performance 
of the inversion algorithm. The shape of the SFR is preserved, and the integrals vary by less than 4\%.
\subsubsection{IMF exponent}
%
%
%
%
%
%
Figure \ref{fig:imf} shows the effect of
changes in the IMF power law exponent of $\pm0.3$. This leads to an over- and under-estimation of the total integrated SFR of about 20\%,
but does not significantly affect the overall shape of the recovered function.
The explanation for this is that variations in the slope of the IMF change the fraction of high mass stars that the algorithm forms.
For a given SFR model, a flatter IMF ($\Delta \imf = +0.3$) will result in a greater number of WDs at all ages, due to the increased
fraction of high mass MS stars with short lifetimes. The magnitude of the SFR will then be reduced by the inversion algorithm in 
order to match the observed WDLF and compensate for the overproduction of WDs, leading to an overall reduction in the SFR that 
is roughly independent of age.
\subsubsection{Metallicity}
%
%
%
%
%
%
The effect of variations in the progenitor metallicity is similar to that of variations in the IMF exponent.
At constant stellar mass, lower metallicity results in shorter main sequence lifetimes, so a reduction in the metallicity parameter
$Z$ results in overproduction of WDs and an overall suppression in the recovered SFR in order to match the observed WDLF. In
Fig. \ref{fig:z}, the true value of $Z$ is 0.017. Reducing to $Z=0.008$ causes the integrated SFR to be underestimated by around 10\%, and
increasing to $Z=0.040$ leads to an overestimation of around 20\%, while in both cases the overall shape of the SFR is preserved. This
assumes of course that the metallicity is independent of time; the existance of an age-metallicity relation would cause the shape 
of the SFR to be incorrectly estimated as well as the normalisation.
\subsubsection{WD Cooling Models}
\label{diff_cooling}
%
%
%
%
%
%
%
Figure \ref{fig:cooling} demonstrates the effect of using different sets of WD cooling sequences. In these tests, the F01 models were
used to generate the observed WDLF, which was then inverted using the S10 and S10p models.
Clearly, the choice of WD cooling models is a significant source
of error in the WDLF inversion algorithm, and for all SFR models this has the largest effect on the integrity of the solution.
Although the integrated SFR is quite well preserved (within a few percent in these tests), the shape of the SFR solution is quite
severely compromised beyond $\sim$2 Gyr in the past, with broad peaks in the SFR shifted significantly (the peak at $\sim$3 Gyr)
or lost entirely (the peak at $\sim$9 Gyr). These effects can be explained as follows.

The loss of the peak at $\sim$9 Gyr in the S10 and S10p solutions is due to the fact that these cooling sequences cool \textit{slower}
than the F01 sequences at faint magnitudes (equivalently, long cooling times). This is important because
when WDs cool slower at certain magnitudes, they traverse the WDLF bins slower and tend to `pile up', leading to larger number
densities. Figure \ref{fig:WDs} demonstrates this differential cooling for
0.6$\msolar$ WDs; the effect is stronger for higher mass WDs, which is significant because these are over-represented at faint 
magnitudes. Because the F01 sequences were used to generate the synthetic WDLF in these tests, when the S10 and S10p sequences are
used to invert it, the recovered SFR is suppressed at old times in order to match the observed WD density in the faint magnitude bins.
We can therefore predict that the S10 and S10p sequences will always recover a lower SFR at old times ($\gtrsim$8 Gyr) than the F01 sequences.

In addition, the inclusion of phase separation effects in the S10p cooling sequences, which slows the cooling of WDs relative to the
S10 sequences at ages greater than $\sim$2 Gyr, has the effect of shifting features in the S10p SFR solution to
older times. This effect is analagous to that described above, and explains the shift in SFR features at more recent times.
\section{The Solar Neighbourhood Star Formation History}
We are now ready to apply the inversion algorithm to a selection of recent determinations of the Solar neighbourhood WDLF.
In these tests, we have fixed the IMF exponent, initial-final mass relation, metallicity and $\alpha$ at the values
listed in Table \ref{tab:params}, which are reasonable for disk stars. The algorithm is not too sensitive to the
chosen values for these parameters. However, we have repeated the inversion for each set of WD cooling models, 
as the solution is likely to be highly cooling model dependent.

Note that the SFR and WDLF models that are generated during the inversion process are dimensionless in spatial density
(their units are just $\mathrm{N} \mathrm{yr}^{-1}$ and $\mathrm{N}  \mathrm{M}_{\mathrm{bol}}^{-1}$), and will be automatically
calibrated to whatever spatial density units the observed WDLF has (normally pc$^{-3}$).
In all cases, the inverted SFR is only for stars more massive than 0.6$\msolar$.
\subsection{The Sloan Digital Sky Survey}
%
%
%
%
%
%
%
%
%
%
The SDSS has produced some of the deepest and cleanest WDLFs in recent years using a variety of different survey techniques.
The WDLF of \citet{harris2006} (H06) is derived from a catalogue of $\sim$6000 WDs obtained from Data Release 3 using
the reduced proper motion technique, with proper motions obtained by combination with USNO-B data. This survey method does
not work well for intrinsically bright stars (which have on average lower proper motions), and their luminosity function
covers the range $7 < \mbol < 16$. At brigher magnitudes, selection of WDs on colour works well due to the UV excess shown
by these objects, and the WDLF of \citet{krzesinski2009} (K09) covers the range $0 < \mbol < 7$, which in conjunction with the H06 LF
provides constraint on the WDLF over nearly the complete range of luminosity.

Figure \ref{fig:harris_sfr} shows the SFR solution obtained when the concatenated H06 and K09 WDLF is inverted.
Although the results differ quite significantly depending on which set of WD cooling models is used,
all show a certain bimodality in the SFR with broad peaks at $\sim$2--3 Gyr and $\sim$6--10 Gyr in the past.
The shape of the SFR functions places the onset of star formation roughly 8--11 Gyr ago depending on the cooling models.
The differences between the solutions are due to differences in the predicted cooling rates of WDs (see Section \ref{diff_cooling});
essentially, the SFR peak at $\sim$6--10 Gyr is higher for the F01 cooling sequences due the the fact that these cool
faster at faint magnitudes, and the SFR at early times is inflated in order match the observed density of faint WDs at the present day.
If the true WD cooling rates are closer to the S10/S10p sequences, then the size of the SFR peak is overestimated in the F01 solution.
Alternatively, if the true cooling rates are closer to the F01 sequences, then the S10/S10p solutions underestimate the peak.
The shift in features between the S10 and S10p solutions older than $\sim2$ Gyr is due to C/O phase separation effects slowing
the cooling of WDs, which pushes features to older times in the S10p solution.
The integrated SFR for each of the models agrees very well: we find $(13.9\pm0.6)\times10^{-3}$
stars/pc$^{3}$ for the F01 solution, $(14.0\pm0.6)\times10^{-3}$ stars/pc$^{3}$ for the S10 solution
and $(13.8\pm0.6)\times10^{-3}$ stars/pc$^{3}$ for the S10p solution.
\begin{figure}
\centering
\includegraphics[trim=1.5cm 0cm 1.8cm 0, clip=true, width=8cm]{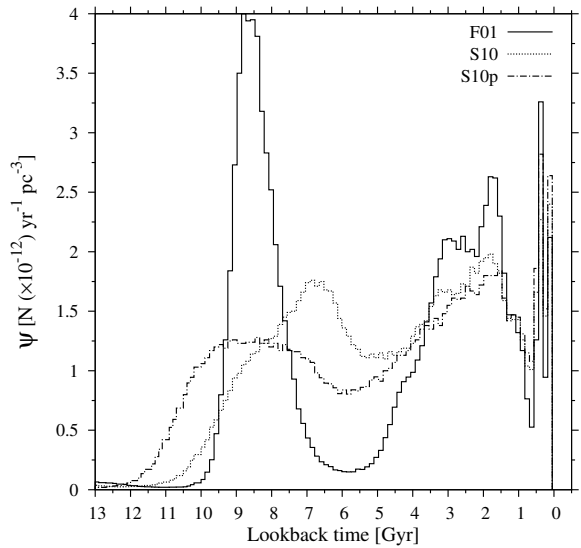}
\caption[]{Star Formation Rate functions obtained by inverting the H06+K09 WDLF. The integrated SFR for each 
of the F01, S10 and S10p solutions is
$(13.9\pm0.6)\times10^{-3}$, $(14.0\pm0.6)\times10^{-3}$ and $(13.8\pm0.6)\times10^{-3}$ stars/pc$^{3}$.
}
\label{fig:harris_sfr}
\end{figure}

Figure \ref{fig:harris_wdlf} shows the final converged WDLF models. These fit the observed WDLF very well, with $\chi^2$ statistics
of 26.4, 53.4 and 44.6 for the F01, S10 and S10p solutions respectively.
\begin{figure}
\centering
\includegraphics[trim=1.5cm 0cm 1.8cm 0, clip=true, width=8cm]{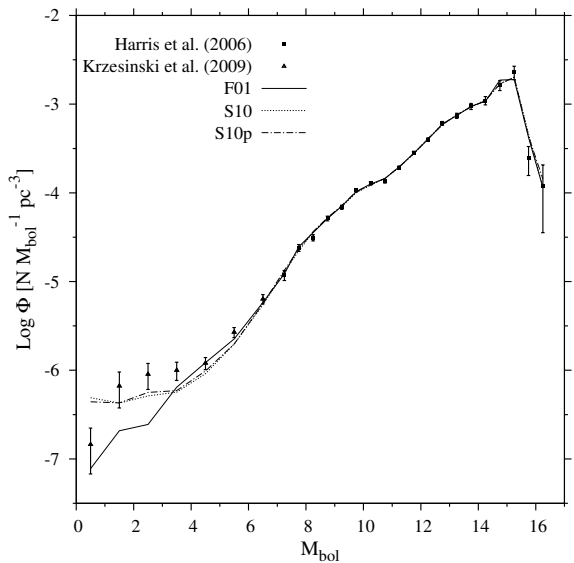}
\caption[]{Converged WDLF solutions using different WD cooling models. The $\chi^2$ statistic for each of the F01, S10 and S10p models
is $26.4$, $53.4$ and $44.6$.
The smaller $\chi^2$ for the F01 solution arises mainly from the brightest WDLF bin: this has a small
observational error relative to the next brightest bins, which is not so apparent due to the log scale.
}
\label{fig:harris_wdlf}
\end{figure}
The residuals of the WDLF models to the data (Fig. \ref{fig:harris_wdlf_ratio}) highlights any systematic errors in the fits.
Generally, we find that the recovered SFR models produce present-day WDLFs that match the observed WDLF very closely.
\begin{figure}
\centering
\includegraphics[trim=1.5cm 0cm 1.8cm 0, clip=true, width=8cm]{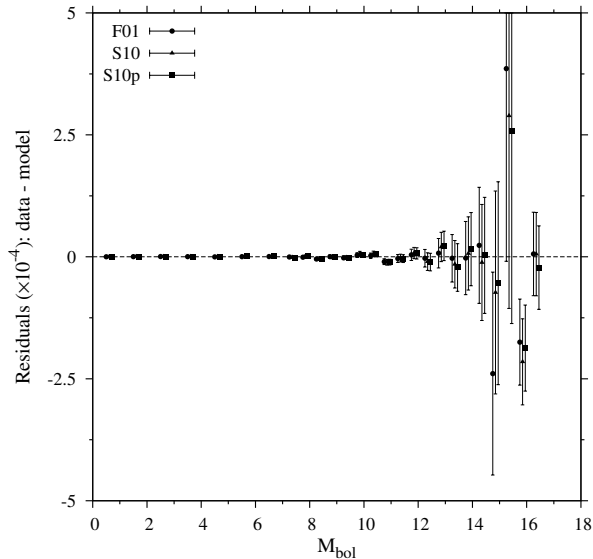}
\caption[]{Residuals for converged WDLF models. The S10 and S10p models have been offset
horizontally by 0.1 and 0.2 magnitudes for clarity. No significant departures from the observed WDLF are seen.}
\label{fig:harris_wdlf_ratio}
\end{figure}
\subsection{The SuperCOSMOS Sky Survey}
%
%
%
%
%
The all-sky SuperCOSMOS Sky Survey (SSS) is based on digitized photographic plates, and is accessible online via an SQL
interface\footnote{\texttt{http://surveys.roe.ac.uk/ssa/}}. The merged source table contains multi epoch photometry in 
three photographic bands $BRI$ and proper motions for nearly two billion objects.
This was used by \citet{rowell2011} (RH11) to measure the WDLF for a sample of around 10,000 WDs,
using the reduced proper motion technique to cover the range $1 < \mbol < 18$.
The RH11 catalogue overlaps with that of H06, but covers a significantly larger area of sky. It is incomplete at around the 50\% level,
but the incompleteness is independent of colour and does not bias the WDLF.
RH11 developed a new method of measuring the WDLF that allowed the different kinematic populations to be resolved in a 
self-consistent way, allowing the thin and thick disk WDLFs to be measured separately for the first time.
In the present work, we use their WDLF derived using the standard $V_{\mathrm{max}}^{-1}$ technique, for a more direct
comparison with the H06 and K09 WDLFs.
In order to achieve a reasonable fit to the RH11 WDLF and ensure smooth convergence of the algorithm, we had to remove both the 
faintest and brightest three points, restricting the WDLF to the range $2.5 < \mbol < 16.5$. The faintest
bins are sparsely populated and contain poorly characterised ultracool WDs, for which the photometric parallaxes are extremely uncertain,
and at the bright end the bins are severely underpopulated leading to highly uncertain data points that cannot be fitted by any SFR.

Figure \ref{fig:rowell_sfr} presents the SFR results obtained by inverting the RH11 WDLF. These again show a clear bimodal form,
with peaks at $\sim$2--3 Gyr and $\sim$6--10 Gyr in the past, depending on the choice of WD cooling models. 
Again, the effects of differential cooling rates between the F01, S10 and S10p sequences explain the differences between the
SFR solutions.
The integrated SFR
is $(8.1\pm0.1)\times10^{-3}$ stars/pc$^{3}$ for the F01 solution, $(8.1\pm0.1)\times10^{-3}$ stars/pc$^{3}$ for the S10 solution
and $(8.0\pm0.1)\times10^{-3}$ stars/pc$^{3}$ for the S10p solution.
These values are around $60$\% of those obtained from the
H06+K09 WDLF, due to the incompleteness present in the RH11 survey.
\begin{figure}
\centering
\includegraphics[trim=1.5cm 0cm 1.8cm 0, clip=true, width=8cm]{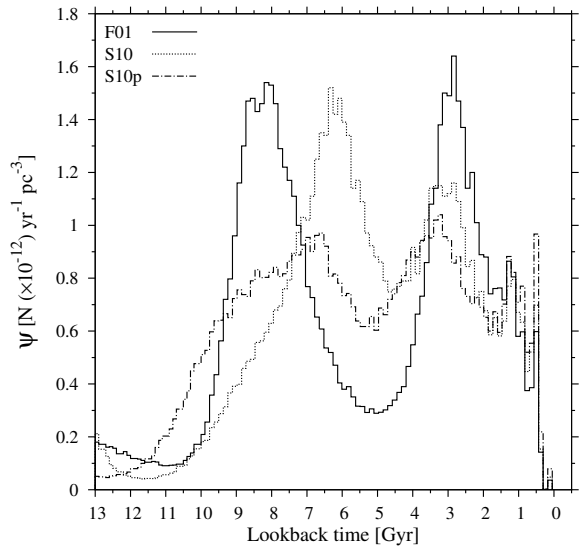}
\caption[]{Star Formation Rate functions obtained by inverting the \citet{rowell2011} WDLF. The integrated SFR for each of the 
F01, S10 and S10p solutions is $8.1\pm0.1\times10^{-3}$, $8.1\pm0.1\times10^{-3}$ and $8.0\pm0.1\times10^{-3}$ stars/pc$^{3}$.}
\label{fig:rowell_sfr}
\end{figure}
Figure \ref{fig:rowell_wdlf} shows the final converged WDLF models. In each case, the fit is not as good as for the H06+K09 WDLF,
achieving $\chi^2$ of 75.9, 63.5 and 62.4 for the F01, S10 and S10p cooling models respectively,
though some of this increase is due to RH11 having two additional data points (28 compared to 26).
\begin{figure}
\centering
\includegraphics[trim=1.5cm 0cm 1.8cm 0, clip=true, width=8cm]{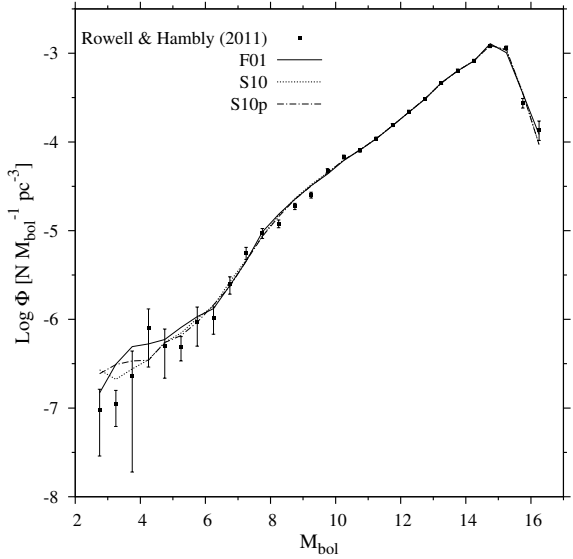}
\caption[]{Converged WDLF solutions using different WD cooling models. The $\chi^2$ statistic for each of the F01, S10 and S10p models
is 75.9, 63.5 and 62.4.}
\label{fig:rowell_wdlf}
\end{figure}
%
In contrast to the H06+K09 WDLF fits, the RH11 WDLF shows some minor deviation in the residuals over the range
$8 < \mbol < 11$ that none of the models have been able to fit (see Fig. \ref{fig:rowell_wdlf_ratio}, inset).
\begin{figure}
\centering
\includegraphics[trim=1.5cm 0cm 1.8cm 0, clip=true, width=8cm]{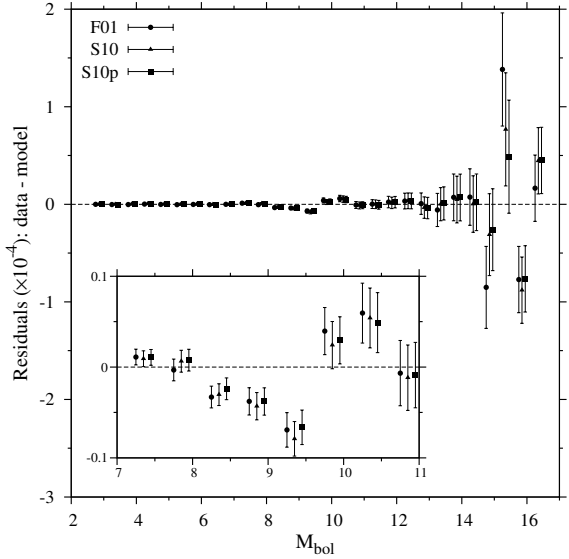}
\caption[]{Residuals for converged WDLF models. The S10 and S10p models have been offset
horizontally by 0.1 and 0.2 magnitudes for clarity. A marginal excess is seen over the range $8 < \mbol < 11$, where none
of the models have been able to adequately fit the data.}
\label{fig:rowell_wdlf_ratio}
\end{figure}
%
%
%
%
%
\subsection{Comparison to other studies}
As far as we are aware, this is the first study to use WDs to probe in detail the time varying SFR of the Galactic disk, and the most directly
comparable work has been done using MS stars; specifically, the colour-magnitude diagram (CMD) for MS stars in the Hipparcos catalogue.

%
\citet{hernandez2000} use a non-parametric Bayesian approach to derive the SFR by inverting the CMD. Their technique provides a high resolution
of $\sim$50 Myr on the recovered SFR, but requires a volume complete sample of stars that restricts their results to the last 3 Gyr. Over this range,
they derive a Solar neighbourhood SFR with an oscillation of period $\sim$0.5 Gyr, superposed on a weak constant rate. Our results hint at
such an oscillation at $t<1.5$ Gyr, although this may just be inversion noise, and the lack of resolution in our SFR prevents us from detecting 
any feature like this further in the past. Also, the method of \citet{hernandez2000} enforced zero SFR at the present day so there may be a risk of
artefacts at early times.

\citet{vergely2002} use a similar Bayesian inversion method, but without the requirement of a volume limited sample, allowing them
to use a much larger number of stars and probe the full star formation history of the disk. They simultaneously fit the SFR and the
age-metallicity relation, obtaining a column-integrated SFR that shows a peak at around $2$ Gyr followed by a smooth decline which levels off
to a constant SFR older than $\sim$5 Gyr. Their results are plotted in Fig. \ref{fig:vergely} over our own results obtained by inverting
the H06+K09 WDLF using the S10p cooling models. Both of the functions are normalised to unity; due to the different
units we compare only the shape of the SFR.
%
%
\begin{figure}
\centering
\includegraphics[trim=0cm 0cm 0cm 0, clip=true, width=8cm]{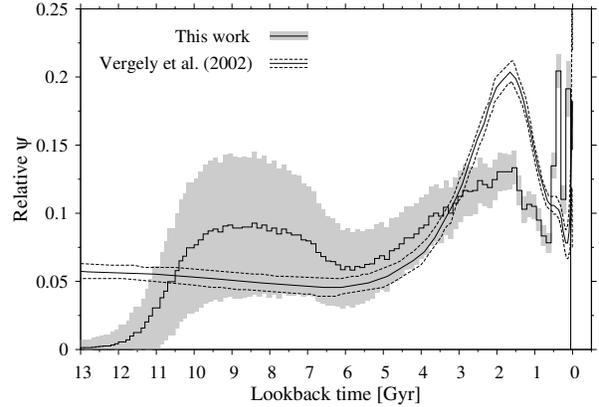}
\caption[]{The results of \citet{vergely2002} compared to our own. No secondary peak in star formation is observed at older times.}
\label{fig:vergely}
\end{figure}

\citet{cignoni2006} use a different technique that combines Bayesian reconstruction of the noise-free Hipparcos CMD with a maximum likelihood
fitting technique for their model CMDs. They again use a volume complete sample, but with a fainter magnitude limit than \citet{hernandez2000}
allowing them to constrain the SFR over the last $12$ Gyr with a time resolution varying from 0.5 to 2.0 Gyr.
Their results are plotted in Fig. \ref{fig:cignoni} along with our own; both functions are again normalised.
They obtain a similar SFR to that of \citet{vergely2002}, with a peak at around 2--3 Gyr followed by a gradual decline to older times, but with
a secondary peak at 10--12 Gyr.
\begin{figure}
\centering
\includegraphics[trim=0cm 0cm 0cm 0, clip=true, width=8cm]{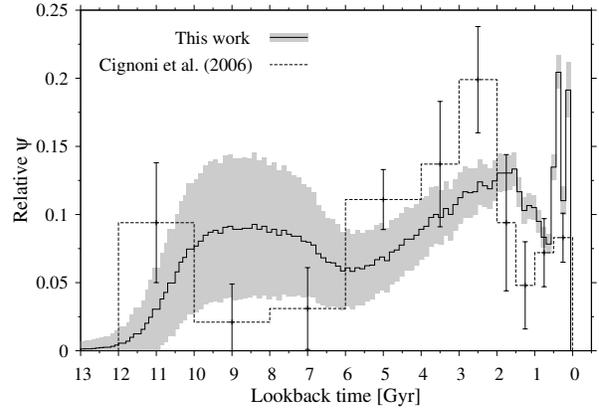}
\caption[]{The results of \citet{cignoni2006} compared to our own. Both studies find a bimodal SFR with a secondary peak at ancient times.}
\label{fig:cignoni}
\end{figure}

The results of both \citet{vergely2002} and \citet{cignoni2006} agree with our own finding of a broad peak in star formation at $\sim$2--3 Gyr ago.
Neither predicts the strong recent bursts in star formation that we obtain from the H06+K09 WDLF, though this is not seen in the 
RH11 WDLF results and may be due to noise.
The secondary peak in star formation at ancient times that is seen in all of our results is not present in the Vergely et al SFR. A similar
feature \emph{is} seen in the Cignoni et al SFR, though the peak is shifted by around $\sim$2--3 Gyr so a positive identification is difficult.
We note that both studies use different values of the IMF slope ($-3.0$ and $-2.35$ respectively, compared to $-2.3$ in the present work), 
though this is not expected to change the shape of the recovered SFR much. Both also use an age-metallicity relation (AMR), the existence of 
which \emph{is} expected to change the shape of our own SFR. The total variation in metallicity in their AMRs is roughly the same
as the range of metallicities considered during testing of our algorithm (though we only use constant metallicity models), 
and the variation in SFR present in Fig. \ref{fig:z} therefore corresponds roughly to the additional uncertainty in the SFR solution 
were such an AMR true. This is not enough to explain the secondary peak seen in our results, so this would seem to be a real feature.
\section{Conclusions}
In this paper, we have presented an algorithm for use in inverting the white dwarf luminosity function to obtain the time
varying star formation history of the host stellar population. We have verified the performance and sensitivity to noise and
various parameters by analysis with synthetic data, and applied the algorithm to two recent independent measurements of the Solar
neighbourhood WDLF. The SFR in the Solar neighbourhood appears to be characterised by
a bimodal distribution with broad peaks at 2--3 Gyr and 7--9 Gyr in the past,
separated by a significant lull. The onset of star formation occurs around 8--10 Gyr ago. These broad results are consistent across
both data sets and are independent of the WD cooling models used.
However, the finer details of the SFR, such as the relative size of the peaks and lull and the precise timing of the various features,
are highly dependent on the choice of WD cooling sequences.

The model WDLFs that the algorithm computes match the observations very well, and we find no systematic deviation that might
indicate additional sources of WD cooling that are unaccounted for in the models. The marginal feature seen at $8 < \mbol < 11$
in the RH11 WDLF is not observed in the H06 WDLF; because the latter uses higher quality SDSS photometry we conclude that
this feature is likely an artefact in the RH11 data. However, we stress that this is not 
evidence \emph{against} the existence of additional cooling processes, rather that differences between simple WDLF models
and the observed WDLF cannot be reliably interpreted as evidence for additional cooling processes
(at least at magnitudes fainter than $\mbol\sim10$) without considering the full time varying star formation history.
Note that it may be possible to use a similar algorithm to measure directly the cooling rates of WDs, if the SFR can be
constrained from other studies.

In principle the algorithm can be applied to any population for which the WDLF can be measured, for example the
spheroid and nearby clusters, although at present the algorithm does not consider binary stars or more exotic 
objects such as He or O/Ne core WDs. This may be a direction for future development work.

\bibliographystyle{mn2e}
\bibliography{references}

\label{lastpage}

\end{document}